\documentclass[superscriptaddress,byrevtex,nofootinbib]{revtex4}
\newcommand{\jj}{$J/\Psi$}

\newcommand{\ie}{{\em i.e.}}
\newcommand{\eg}{{\em e.g.}}
\newcommand{\etal}{{\it et al.}}
\usepackage[dvips]{graphics,graphicx,epsfig}
\usepackage{subfigure}
\usepackage{latexsym}
\begin{document}
\title{Charmonium Suppression and Regeneration from SPS to RHIC}
\date{\today}
\author{L. Grandchamp}
\email{loic@tonic.physics.sunysb.edu}
\affiliation{Department of Physics and Astronomy, State University of
  New York at Stony Brook, Stony Brook, NY 11794-3800}
\affiliation{Institut de Physique Nucl\'eaire de Lyon, 43 Bd du 11
  Novembre, 69622 Villeurbanne cedex}

\author{R. Rapp}
\affiliation{Department of Physics and Astronomy, State University of
  New York at Stony Brook, Stony Brook, NY 11794-3800}

\begin{abstract}
The production of charmonia is investigated for heavy-ion collisions 
from SPS to RHIC energies. Our approach incorporates 
two sources of \jj\ yield: (i) a direct contribution  
arising from early (hard) parton-parton collisions, subject to 
subsequent nuclear absorption, quark-gluon plasma
and hadronic dissociation, and (ii) statistical production  
at the hadronization transition by coalescence of $c$ and
$\bar{c}$ quarks. Within an expanding thermal fireball framework, 
the model reproduces \jj\ centrality dependencies observed at the
SPS in $Pb$-$Pb$ and $S$-$U$ collisions reasonably well. The study 
of the $\Psi'/\Psi$ ratio at SPS points at the importance of
the hadronic phase for $\Psi'$ interactions, possibly related to effects 
of chiral symmetry restoration.  Predictions are given for the 
centrality dependence of 
the $N_{J/\Psi}/N_{c\bar{c}}$ ratio at full RHIC energy. We also 
calculate the excitation function of this ratio. The latter exhibits 
a characteristic minimum structure signalling the transition from 
the standard \jj\ suppression scenario prevailing at SPS to dominantly 
thermal regeneration at collider energies.
\end{abstract}
\maketitle

\section{Introduction}
The main objective of ultrarelativistic heavy-ion physics, both
experimentally and theoretically, is to create and study hot and dense
matter of strongly interacting constituents. At sufficiently large   
temperatures and densities, the theory of strong
interactions (QCD) predicts a phase transition from the hadronic world
into a chirally symmetric state of free quarks and gluons, the so-called
Quark-Gluon Plasma (QGP). However, the transient nature of this new
state of matter renders its identification very complex. After more 
than a decade of experiments, a wealth of exciting results has emerged from 
the SPS program as well as from recent data taken at the the 
Relativistic Heavy-Ion Collider (RHIC) at Brookhaven. However, as of now, 
no conclusive evidence of QGP formation has been discerned. 
It has rather become clear that the discovery of the QGP would not emerge 
from a single signature but from a combination of several in a comprehensive
scenario.

Among the probes of QGP formation, the \jj\ meson plays a central role. 
Due to its large binding energy 
and rather small size, it is expected to be only mildly affected by 
interactions  in a hadronic environment. On the contrary, as suggested first by
Matsui and Satz~\cite{MS86} in 1986, if a QGP is formed, \jj\ 
formation should be inhibited due to color charge screening by free-moving
quarks and gluons. From a slightly different perspective, a similar
effect can be attributed to parton-induced destruction~\cite{Shu78}.  
In either picture, the \jj\ abundance observed in heavy-ion collisions is 
linked  to properties of the initial hot and/or dense phases of the 
produced matter. However, the identification of the plasma effect 
requires a careful assessment of other mechanisms for \jj\ 
destruction in the course of a heavy-ion reaction. Most notably,   
absorption on high-energetic nucleons at early times induces 
a substantial suppression of typically 60\% in central collisions.
Although its dependence has been thoroughly investigated experimentally
in both proton and light-ion induced reactions with various 
targets~\cite{na38-91,na38-98,na38-99a,na50-02b} 
theoretical uncertainties in the application to the heavy-projectile case
persist~\cite{QVZ02}.  
Furthermore, despite the naive expectation given above, the current 
status in the description of inelastic charmonium scattering on 
``comoving'' (secondary) hadrons in the late stages of the collision 
is not satisfactorily under control.  
This renders the identification of the effect of the quark-gluon
plasma formation on \jj\ abundances more difficult to establish.

Recently, an additional source of \jj\ {\em production} in
heavy-ion collisions has been suggested.
Within thermal model frameworks~\cite{pbm96-99,pbm01,Beca01}, 
which successfully describe light-hadron ratios, 
statistical production of charmonium at the QCD hadronization transition 
has been evaluated~\cite{GG99,BS00a,BS00b,GKSG01}.   
Gazdzicki and Gorenstein found~\cite{GG99} that statistical production 
alone can account for the observed centrality dependence of \jj\ yields 
at full SPS energy, thereby deducing a hadronization temperature consistent 
with standard chemical freezeout. However, no reference was made to  
a dynamical origin for  $c\bar c$ creation, nor to open-charm states.     
Braun-Munzinger and Stachel~\cite{BS00a,BS00b} started from the  
dynamically well justified~\cite{Re00} hypothesis
that at SPS energies all $c\bar{c}$ pairs are exclusively created in  
primordial (hard) parton-parton collisions. Open- and hidden-charm
hadrons are then populated at the hadronization transition according
to their thermal weight including a charm-quark fugacity to match
the primordial $c\bar c$ abundance. With the latter taken from
extrapolations of  $N$-$N$ collisions, a substantial fraction of
\jj\ mesons is found. In particular, the $\Psi'$ over $\Psi$ ratio, 
which in $Pb$-$Pb$ reactions has been observed to rapidly approach its 
thermal value for $N_{part}\ge 100$,  is well described. 
Along similar lines, Gorenstein \etal~\cite{GKSG01} aim at explaining    
the \jj\ yields at SPS  solely in terms of statistical production.  
They conclude that an open-charm enhancement factor of up to $\sim$~6 
in central $Pb$-$Pb$ collisions (w.r.t. the standard value inferred from 
$N$-$N$ collisions) is required to optimally reproduce the NA50 
data~\cite{KG01,KGSG02}. Finally, Thews \etal~\cite{TSR00} 
assessed $c\bar c$ coalescence in an expanding QGP fireball by   
solving rate equations for   $J/\Psi +g \leftrightarrow c \bar c$ 
reactions (gluon-induced ``photo''-dissociation and its reverse
reaction), with the \jj\ binding energy assumed to be at its vacuum
value at all times in the evolution. 

All thermal approaches share the common feature that, as   
first pointed out in Ref.~\cite{BS00a}, at RHIC energies a copious 
production of charm-quark pairs ($N_{c\bar c}\simeq$~10-20 in central 
$Au$-$Au$ collisions) implies a much enhanced charmomium
yield as compared to early (hard) production coupled with 
nuclear and QGP suppression mechanisms.  
However, at SPS energies, expected plasma lifetimes 
of $\sim$~1-2 fm/c, together with initial temperatures 
below $\sim$~250~MeV, may not suffice for a (close to) complete 
destruction of primordially produced of \jj\ mesons.  
At the same time, a significant increase in open-charm production 
is not easily justified.  
This led us to propose a two-component model~\cite{GR01} 
for charmonium production in heavy-ion collisions. It incorporates both  
a primordial yield  (subject to subsequent dissociation) 
as well as  a thermal contribution from statistical recombination
of $c$ and $\bar{c}$ quarks at the hadronization transition. 
Both components are evaluated within a common thermal fireball framework 
which is consistent with measured hadro-chemistry, expansion
dynamics, and has also been successfully employed to describe 
electromagnetic observables at SPS energies~\cite{RW00}. 
Another important feature is that we refrain from  
invoking any ``anomalous'' open-charm enhancement. 

In the present article, we expand upon our previous analysis in several
respects: First, and most importantly, we give a detailed account 
of centrality dependencies for $J/\Psi$ yields in both $S$-$U$ and 
$Pb$-$Pb$ systems at SPS,
as well as predictions for $Au$-$Au$ at full RHIC energy. The relevance 
of hadronic dissociation is assessed, especially in the context of the
$\Psi'/\Psi$ ratio measured at SPS. Higher charmonium states relevant 
for feeddown contributions to the $J/\Psi$ are included on an 
equal footing. Effects of incomplete charm-quark thermalization, which 
affect thermal recombination, are incorporated on
a phenomenological basis. We also lay out 
further steps to investigate problems that will not be satisfactorily
addressed here.    
 
The paper is organized as follows: In Sect.~\ref{sec:nucsupp}, we
recall basic elements of nuclear absorption including normalization
issues when comparing to NA38 and NA50 data. 
In Sect.~\ref{sec:supp}, we present our calculations for the charmonium  
dissociation processes, \ie, parton-induced destruction in a QGP    
as well as inelastic hadronic \jj\ interactions using effective Lagrangians. 
Pertinent survival probabilites are obtained from a convolution of the 
destruction rates over the space-time history of a collision modeled 
within an expanding fireball. Sect.~\ref{sec:thermal}
is devoted to the description of statistical charmonium formation    
at the hadronization transition including exact charm conservation
constraints. In Sect.~\ref{sec:sps}, we combine the two sources of 
charmonia to investigate the \jj\ centrality dependence and $\Psi'/\Psi$ 
ratio at SPS for both  $S$(200 AGeV)-$U$ and $Pb$(158 AGeV)-$Pb$ systems.
In Sect.~\ref{sec:rhic}, we give predictions for the ratio
$N_{J/\Psi}/N_{c\bar{c}}$ as a function of centrality at RHIC, and 
discuss the excitation function of this ratio from SPS to RHIC energies. 
We summarize and conclude in Sect.~\ref{sec:concl},  and indicate  
directions for future improvements.

\section{Nuclear absorption of charmonia}
\label{sec:nucsupp}
A suppressed abundance of \jj\ mesons relative to expectations from 
$N$-$N$ collisions  was first observed in proton-nucleus ($p$-$A$)
and light-ion induced collisions. In these reactions, one does not expect
noticeable effects from secondary particles. Indeed, rather extensive 
experimental studies \cite{na38-91,na38-98,na38-99a,na50-02b} have
established that the suppression can be  
understood as the absorption of a ``pre-resonance'' $c \bar c$ 
state in the (normal) nuclear medium of target and projectile 
nuclei. This in particular accounts for the
observation that different charmonium states such as 
\jj\ and $\Psi'$ follow very similar absorption patterns, 
despite their different bound state properties (\eg, binding energy or
size). 

Let us briefly summarize the main elements of this ``nuclear absorption''.   
In the Glauber model of $A$-$B$ collisions, the probability
for the pre-resonance state to be absorbed on its way through
the nucleus is given by~\cite{KLNS96}
\begin{eqnarray}
  {\cal S}_{nuc}(b,\sigma_{nuc}) &=&  \frac{1}{T_{AB}(b)} \int\limits_{}^{} 
  d^2s\;dz\;dz'\;\rho_{A}(\vec{s},z)\;\rho_{B}(\vec{b}-\vec{s},z') \nonumber 
\\ 
  & \times & \exp \left\{-(A-1)\int_z^{\infty}
    dz_A \rho_{A}(\vec{s},z_A) \sigma_{nuc}\right\} \times 
 \exp \left\{-(B-1)\int_{z'}^{\infty}dz_B\rho_{B}
    (\vec{b}-\vec{s},z_B)\sigma_{nuc} \right \}  
\end{eqnarray}
where $\rho_{A}(\vec{s},z)$ describes nuclear density profiles
(taken from Ref.~\cite{VJ74}), $\vec{s}$ the
position of the $c\bar c$ production point in the transverse plane, and
$b$ the impact parameter. The coordinates $z$ and $z'$ specify the 
positions within nucleus $A$ and $B$, respectively, along the collision 
axis. $T_{AB}(b)$ is the usual nuclear overlap function of the two
colliding nuclei. The pre-resonance ``absorption'' cross section 
$\sigma_{nuc}$ is treated as a free parameter being adjusted to  
experimental data.

In the NA38/NA50 experiments, from which all data shown  
below are taken, the main measure for the centrality of a nuclear 
collisions is given by the transverse energy $E_T$ detected in the 
calorimeter.  The impact parameter $b$ is commonly related to this observable 
through the wounded nucleon model, which, upon inclusion of fluctuations 
in $E_T$ at fixed $b$, provides a correlation function 
\begin{equation}
  {\cal P}_{AB}(E_T,b) = \frac{1}{\sqrt{2\pi q^2 a
      N_{part}(b)}}\exp\left
    \{-\frac{[E_T-qN_{part}(b)]^2}{2q^2aN_{part}(b)}\right\} \ .
\label{eq:PAB}
\end{equation}
Here, $q$ is a proportionality factor between $E_T$   
and the number of participant, $E_T(b)=qN_{part}$, which depends on the
specific experimental settings, and $a$ is an
empirical parameter characterizing the magnitude of the fluctuations.
The absorption of charmonium in
nuclear matter as a function of $E_T$ then follows as  
\begin{equation}
  \frac{B_{\mu\mu}\sigma_{J/\Psi}}{\sigma_{DY}} =
  \frac{B_{\mu\mu}\sigma_{J/\Psi}^{pp}}{\sigma_{DY}^{pp}}
  \frac{\int d^2b\;{\cal P}_{AB}(E_T,b){\cal
      S}_{nuc}(b,\sigma_{nuc})T_{AB}(b)}{\int d^2b\;{\cal
      P}_{AB}(E_T,b)T_{AB}(b)} 
\label{eq:JDYrat}
\end{equation}
where $B_{\mu\mu}$ is the branching ratio for $J/\Psi\to \mu^+\mu^-$.  
The pre-factor $\sigma_{J/\Psi}^{pp}/\sigma_{DY}^{pp}$ is
not accurately known, and depends on the energy of the collision. 

\begin{figure}[!h]
\centering
\mbox{\subfigure{\epsfig{figure=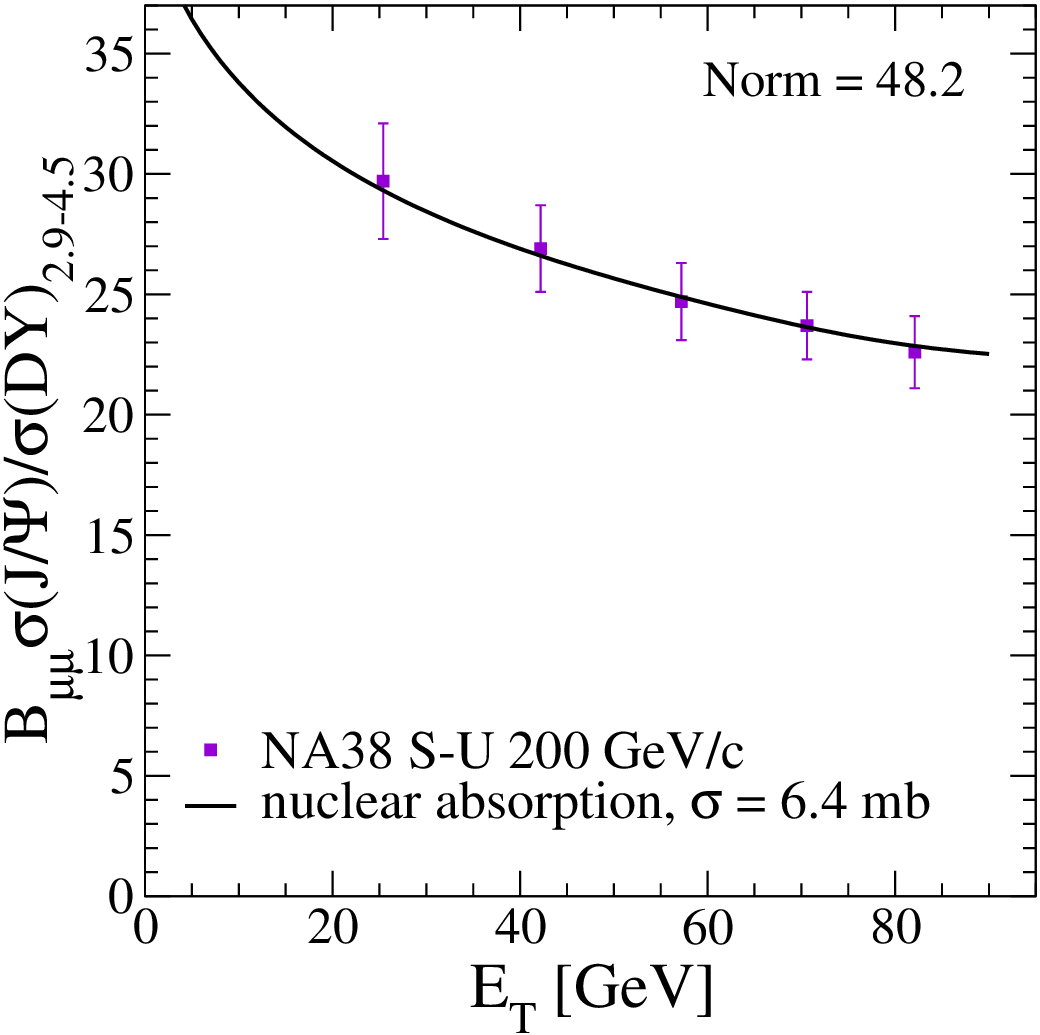,width=0.385\textwidth,clip=}}
  \quad \subfigure{\epsfig{figure=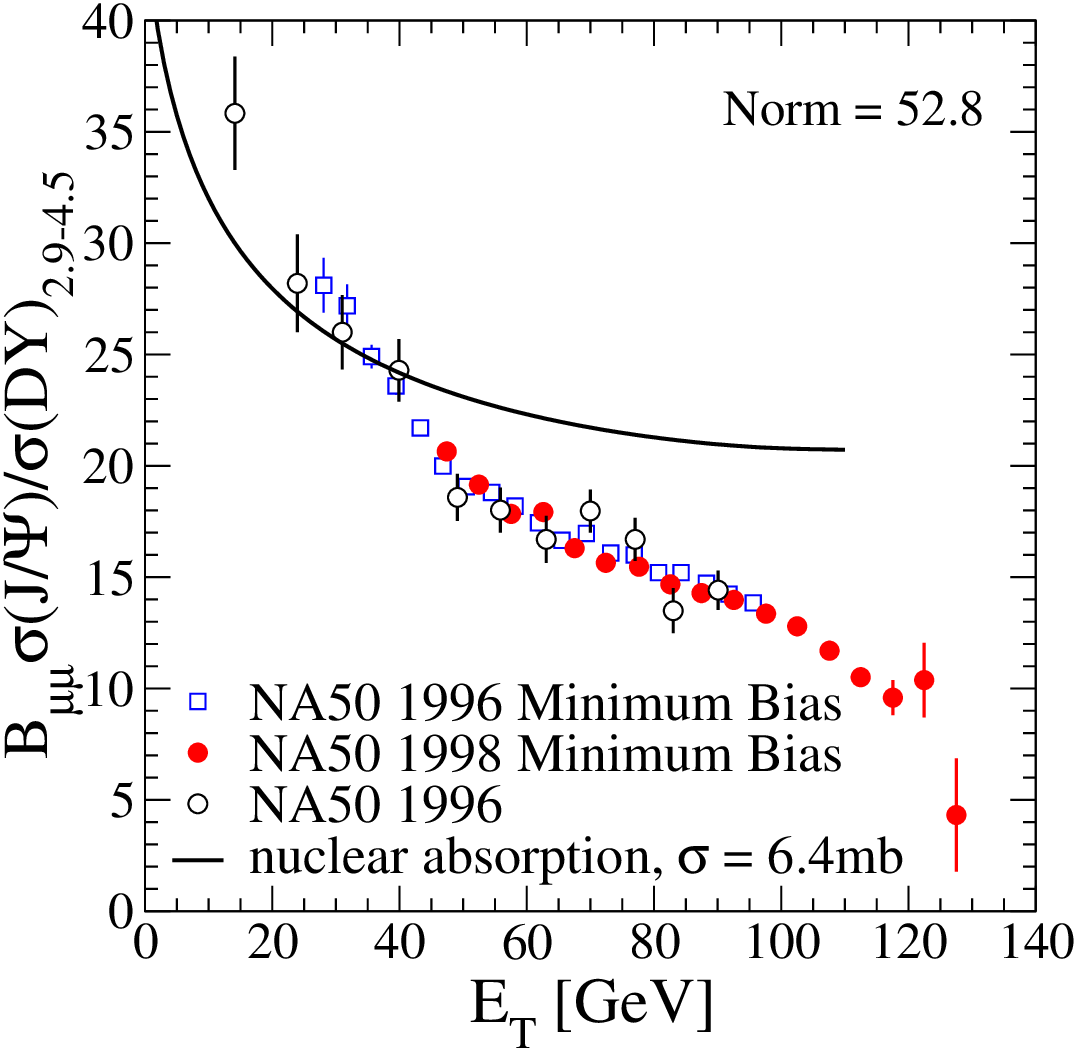,width=0.4\textwidth,clip=}}}
\caption{Nuclear absorption calculations for the $S$(200 AGeV)-$U$ (left 
panel) and $Pb$(158 AGeV)-$Pb$ (right panel) 
systems at the CERN-SPS, compared to data from NA38~\cite{na38-99a} and 
 NA50~\cite{na50-02a}, respectively. The
  normalization factor $B_{\mu\mu}\sigma_{pp}^{J/\Psi}/\sigma_{pp}^{DY}$ 
 has been fixed at 48.2 (52.8) for the $S$-$U$ ($Pb$-$Pb$) system.} 
\label{fig:nuc_abs}
\end{figure}

At SPS energies, we will concentrate on the two collision systems 
$S$(200 AGeV)-$U$ and $Pb$(158AGeV)-$Pb$. 
For $q$ and $a$ we use the values reported by the experiments,
\ie, $q=0.275$ (0.72) and $a=1.27$ (1.56) for $Pb$-$Pb$ ($S$-$U$). 
Together with an absorption cross section of $\sigma_{nuc}=6.4$~mb
and normalization factors of 52.8 for $Pb$-$Pb$ and 48.2 for $S$-$U$, 
our results coincide with nuclear absorption studies of NA38 and NA50, 
cf.~Fig. \ref{fig:nuc_abs}. 
Whereas the $S$-$U$ data are well accounted for,
the $Pb$-$Pb$ system departs from nuclear
absorption starting at $E_T \simeq 40$~GeV, constituting
``anomalous'' \jj\ suppression~\cite{na50-97a,na50-00a,na50-01b,na50-02a}.

\section{$J/\Psi$ dissociation in heavy-ion collisions}
\label{sec:supp}
The deviation of the \jj\ yield from nuclear absorption systematics in 
$Pb$(158 AGeV)-$Pb$ has triggered extensive theoretical analyses.
The responsible underlying mechanisms are still a matter of debate, 
ranging from destruction in a QGP and/or hadron gas to thermal production
at $T_c$. As stressed above, this work attempts a comprehensive treatment  
of all three of these aspects within a minimal set of assumptions, which, 
in particular, is based on a thermal description of the collision dynamics  
including Quark-Gluon Plasma formation if the initial conditions
are energetic enough. In this section, we compute  
\jj\ survival probabilities for both the plasma and 
hadronic phases of heavy-ion reactions.

\subsection{Quark-Gluon Plasma}
\label{sec:qgpsupp}
\jj\ dissociation in a thermalized QGP has been studied in both static 
and dynamical frameworks. Within the former, one typically evaluates the 
screening of the heavy quark potential by color charges, whereas the latter 
involves inelastic parton collisions using, \eg, the QCD analogue of 
photo-dissociation, $g + J/\Psi \rightarrow c + \bar{c}$~\cite{PB79}. 

Here we adopt a dynamical approach, accounting however for reduced 
charmonium binding energies as extracted from a Schr\"odinger equation
for $c\bar{c}$ bound states in a screened heavy-quark potential~\cite{KMS88}.  
Under conditions relevant to the initial stage of heavy-ion
collisions, the binding energy of the \jj\ is strongly reduced
with respect to its vacuum value, which is also borne out of recent 
lattice gauge calculations~\cite{pds01}. 
Color-screening is characterized by the electric  Debye mass 
$\mu_D$ which we estimate to leading order in perturbation theory, 
$\mu_D \sim gT$. Since, strictly speaking, perturbation theory is not 
really applicable under the moderate plasma temperatures expected
even at RHIC energies, we regard the strong coupling constant
as an effective parameter to be adjusted to the \jj\ data at SPS. 
{\it E.g.}, with 
a typical $g \simeq 1.7$, the \jj\ binding energy at $T=170$~MeV is
$E_{diss}=250$~MeV decreasing to $E_{diss}=100$~MeV at $T=230$~MeV and 
vanishing around $T \simeq 360$~MeV. For such small  
binding energies, the photo-dissociation process becomes  
inefficient due to unfavorable break-up kinematics. 
For a loosely bound charmonium state, inelastic parton scattering, 
$g(q, \bar{q}) + J/\Psi \rightarrow g(q, \bar{q}) + c + \bar{c}$,  
turns out to be a more important mechanism~\cite{GR01}. 
The respective cross sections are evaluated in quasifree approximation
using leading-order QCD for $gc \rightarrow gc$ ($qc \rightarrow
qc$)~\cite{Com79}. The thermal dissociation rate is then obtained via 
\begin{equation}
\label{eq:rate}
\Gamma_{diss} = \sum\limits_{i=q,g}^{}
\int\limits_{k_{min}}^{\infty}\frac{d^3k}{(2\pi)^3}f^i(k,T)\sigma_{diss}(s)
\end{equation} 
where $k_{min}$ denotes the minimal on-shell momentum of a quark or
gluon from the heat bath necessary to dissolve an in-medium charmonium
bound state into a free $c\bar{c}$ pair.
We include thermal quasiparticle masses for light quarks and 
gluons~\cite{Bel96}, 
\begin{equation}
  \label{eq:thmass}
  m_{u,d}^2 = \frac{g^2T^2}{6}, \qquad m_s^2 = m_0^2 +
  \frac{g^2T^2}{6}, \qquad m_g^2  = \frac{g^2T^2}{2} \ . 
\end{equation}
This formalism straightforwardly applies to $\Psi'$ and $\chi$ states 
as well.  From the Schr\"odinger equation, it follows~\cite{KMS88}
that their binding energy vanishes at or even below $T_c$, which is also
found in recent lattice calculations~\cite{pds01}. Therefore, the pertinent  
QGP dissociation rate is given by Eq.~(\ref{eq:rate}) with 
$E_{diss}=0$. The picture we have in mind here is similar in spirit 
to that underlying nuclear absorption: a pre-resonance state,
characterized by a correlation induced in the primordial $N$-$N$ collision  
(\eg, comoving $c$ and $\bar c$ quarks), is only destroyed if it actually 
undergoes an inelastic interaction with the surrounding matter. 

\begin{figure}[htb]
\includegraphics[width=0.45\textwidth,clip=]{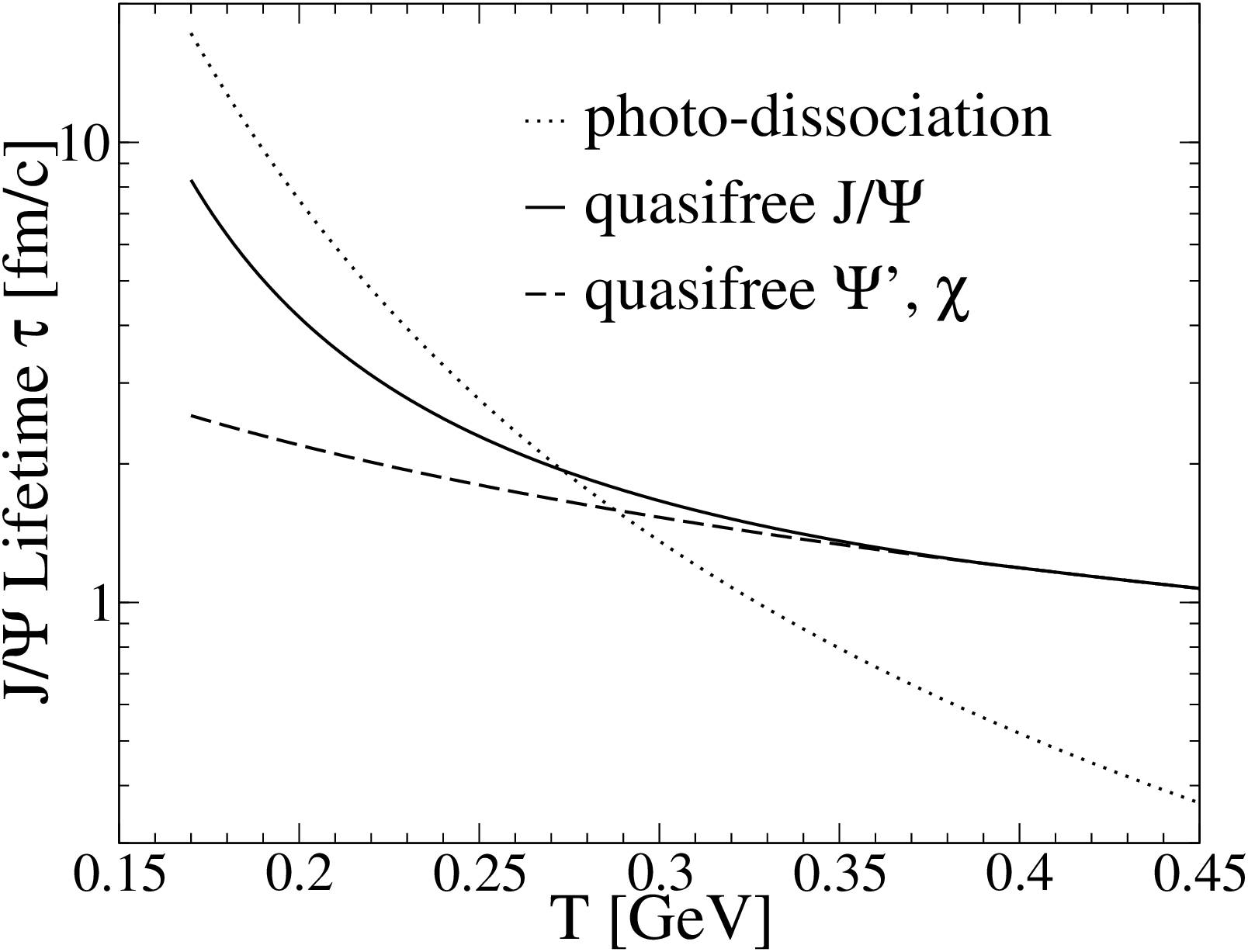} 
\caption{Dissociation times for a $J/\Psi$ ($\Psi'$, $\chi$) in a QGP as a
  function of temperature. The full (dashed) curve corresponds to the 
  leading-order QCD process for quasifree $g,q+c\rightarrow g,q+c$
  scattering with {\em in-medium} $J/\Psi$ ($\Psi'$, $\chi$) bound
  state energies. The dotted curve results from  
  photo-dissociation, $gJ/\Psi \rightarrow c\bar{c}$, assuming
  the {\em vacuum} dissociation energy.} 
\label{fig:taudiss}
\end{figure}

The resulting charmonium dissociation times, 
$\tau_{diss} = \Gamma_{diss}^{-1}$, are shown in Fig.~\ref{fig:taudiss} 
and, for the \jj, are compared to 
photo-dissociation {\em without} medium effects in the bound state 
energy (as has been employed in the literature before).
For temperatures relevant at SPS, the quasifree process (full line) 
is more efficient in dissolving $J/\Psi$'s than
photo-dissociation (dotted line). At higher temperatures this tendency is
reversed, mostly  due to an increasing Debye mass which suppresses the
$t$-channel gluon-exchange graphs for $g(q,\bar q)+ J/\Psi \to g(q,\bar q) +
c + \bar c$. Close to $T_c$, the vanishing binding energies for $\Psi'$ and 
$\chi$ entail lifetimes which are about a factor of three below the 
one for $J/\Psi$-mesons. 

\subsection{Hadron Gas}
\label{sec:hadsupp}
Subsequent to quark-gluon plasma dissociation, when the system
converts to the hadronic phase, charmonia  
undergo further suppression due to interactions with surrounding 
hadrons. There are a number of theoretical models for  
$J/\Psi + hadron$ processes~\cite{KS94,MBQ95,MM98,Vogt99,Ha00,LK00,BSW01,HG01} 
whose results span an appreciable magnitude. Calculations involving 
excited charmonia, such as $\Psi'$ and $\chi_c$ (which contribute 
to the \jj\ yield via electromagnetic feeddown), are scarce. 
Consequently, the impact of inelastic hadronic scattering is not very well
under control. Most calculations, including recent ones based on
quark exchange models~\cite{BSW01} or on $SU(4)$-symmetric effective
lagrangians~\cite{LK00,HG01}, seem to indicate a rather moderate impact of
the hadronic medium on the \jj. 

\subsubsection{Absorption on light hadrons}
\begin{figure}[htb]
\includegraphics[width=0.45\textwidth,clip=]{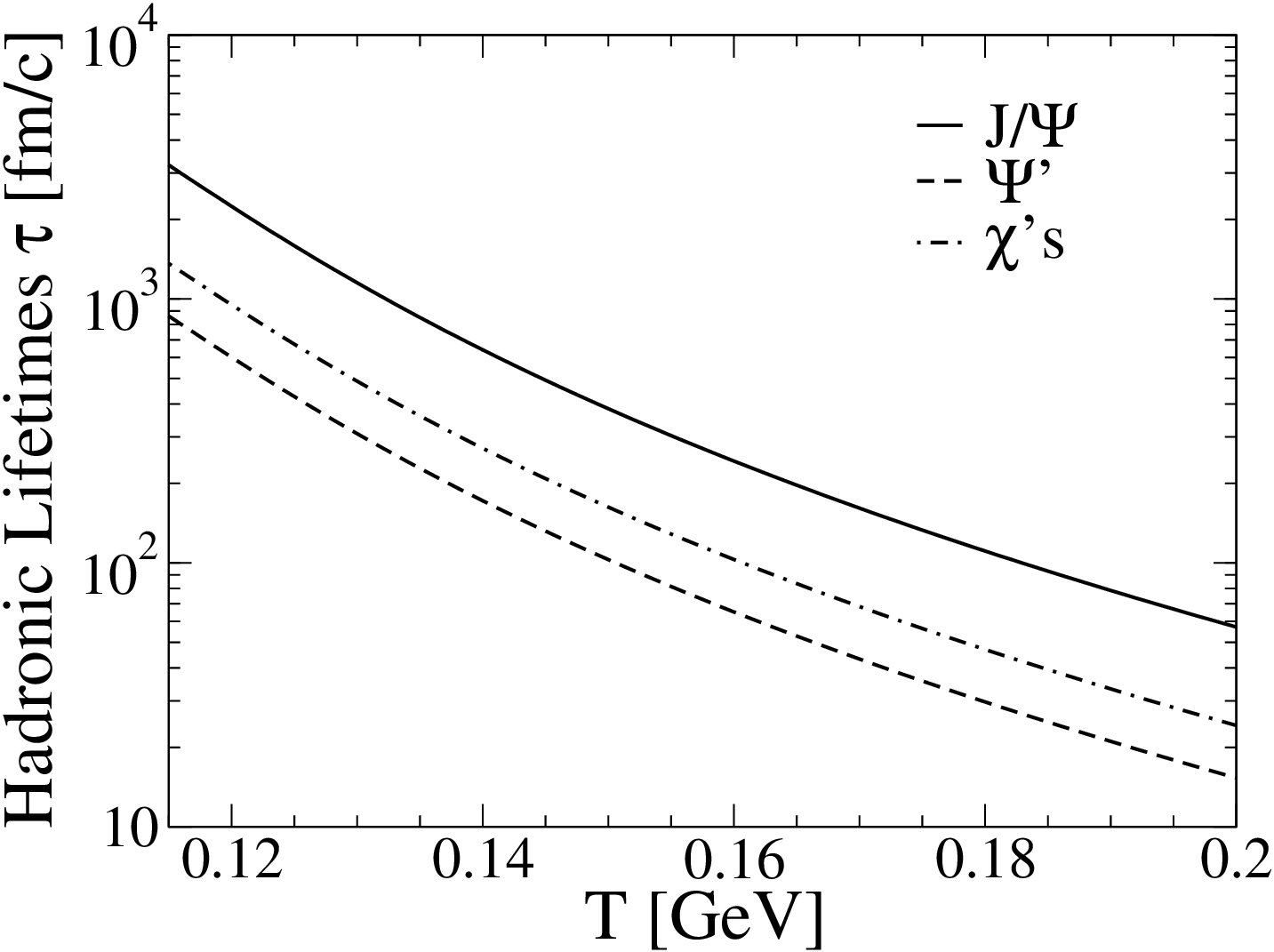} 
\caption{Hadronic lifetimes for \jj\ (solid line), $\Psi'$ 
(dashed line) and $\chi$ (dot-dashed line) based 
on a $SU(4)$ effective lagrangian,
including interactions with pions and rhos using a form factor value
of $\Lambda =1$ GeV. The curves for the $\Psi'$ ($\chi$) are deduced from
the \jj\ one via cross section scaling by $(r_{\Psi'}/r_{J/\Psi})^2$
($(r_{\chi}/r_{J/\Psi})^2$ for the $\chi$).}
\label{fig:tauhad}
\end{figure}

As a baseline calculation, we here reproduce results obtained in 
Refs.~\cite{LK00,HG01} within a $SU(4)$ effective theory. 
The starting point is a $SU(4)$-flavor symmetric
effective lagrangian formulated with 4-by-4 pseudoscalar and vector meson
matrices. Although the $SU(4)$ symmetry is strongly broken by the 
$c$-quark mass, the hope is that symmetry-breaking effects are largely 
accounted for by the hadronic mass matrix. Along these lines  
we compute (inelastic) interactions of the 
\jj\ with pions ($\pi + J/\Psi \rightarrow D + \bar{D}^{\star},
\bar{D} + D^{\star}$) and rhos ($\rho + J/\Psi \rightarrow D +
\bar{D}$, $\rho + J/\Psi \rightarrow D^{\star} + \bar{D}^{\star}$) 
which are the most abundant mesons in the
medium. We employ coupling constants as calibrated by Haglin and
Gale in Ref.~\cite{HG01} based on the $\rho \rightarrow \pi\pi$ decay 
to fix the gauge coupling. The effective hadronic theory is supplemented 
by vertex form factors to simulate finite-size effects. This violates 
gauge invariance, which, however, can be restored by introducing 
appropriate counter terms~\cite{HG01}. 
We have checked for various processes (\eg, $\pi + J/\Psi
\rightarrow D + \bar{D}^{\star}$) that these extra terms induce only
quantitatively minor modifications, which we therefore neglected.   
The corresponding \jj\  lifetime in a $\pi$-$\rho$ gas is 
displayed in Fig.~\ref{fig:tauhad} using (covariant) monopole form factors 
with cutoff $\Lambda=1$~GeV. 

As mentioned above, $\Psi'$ and $\chi$ dissociation rates 
are even more difficult to assess within hadronic model frameworks. 
To obtain an estimate that can be used in our calculations below, 
we assume a geometric scaling of the calculated \jj\ 
cross sections with the squared ratio of the respective 
charmonium radii \cite{KMS88}, $(r_{\Psi'}/r_{J/\Psi})^2$ and
$(r_{\chi}/r_{J/\Psi})^2$, cf. dashed and dot-dashed lines in 
Fig.~\ref{fig:tauhad}. It turns out that these estimates agree 
reasonably well with explicit calculations of the corresponding processes
within the constituent quark model approach of Ref.~\cite{BSW01}. 

\subsubsection{Anomalous Processes}
We furthermore investigated the role of so-called anomalous processes 
such as $\pi + J/\Psi \rightarrow \eta_c + \rho$, 
$\pi + J/\Psi \rightarrow \eta_c + b_1$, and 
$\rho + J/\Psi \rightarrow \eta_c + \pi$, suggested in Ref.~\cite{HG01}.
In there, the relevant hadronic coupling constants, 
\eg, $g_{J/\Psi \omega \eta_c}$, have been estimated applying the vector
dominance model (VDM) to the radiative decay $J/\Psi \to \gamma \eta_c$.
We believe, however,  that this procedure leads to a significant 
overprediction of the hadronic coupling, for the following reason. 
For vertices carrying identical quantum number structure, 
namely $J/\Psi \omega \eta$ and $J/\Psi \omega \eta'$, both hadronic
and corresponding radiative decay information is available from 
experiment. One finds~\cite{pdg00},  
\begin{eqnarray}
\Gamma(J/\Psi \rightarrow \omega \eta)/\Gamma_{tot} =1.6\times 10^{-3}
 , & &   
\Gamma(J/\Psi \rightarrow \omega \eta')/\Gamma_{tot}=1.7\times 10^{-4}
\ ,
\nonumber\\
\Gamma(J/\Psi \rightarrow \gamma\eta)/\Gamma_{tot} =8.6\times 10^{-4} 
 , & & 
\Gamma(J/\Psi \rightarrow \gamma\eta')/\Gamma_{tot}=4.3\times 10^{-3}  \ ,  
\nonumber
\end{eqnarray}
\ie, the hadronic branching ratios are comparable  or even below the 
radiative ones. This is in marked contradiction to VDM 
within which the latter are suppressed by a factor $(e/g_\omega)^2$
which is much smaller than the moderate increase in phase space due to the 
final state decay momenta (also note that VDM is more strongly violated with 
increasing mass of the pseudoscalar meson). Thus it appears that VDM cannot 
be applied, and that an accordingly reduced $g_{J/\Psi \omega \eta_c}$ 
coupling renders the pertinent  $t$-channel $\omega$ exchange processes in  
$\pi + J/\Psi \rightarrow \eta_c + \rho$ and 
$\pi + J/\Psi \rightarrow \eta_c + b_1$ negligible. 
We therefore decided not to include anomalous processes in our analysis.

\subsection{\jj\ survival probability in a thermal expansion scenario}
\label{sec:fireball}
The final number of primordial \jj's remaining after plasma and hadronic 
phases requires the convolution of the dissociation rates derived in the 
previous two sections over the space-time history of a given heavy-ion 
collision.  To this end, we model this evolution by a schematic thermal 
fireball expansion~\cite{RW99}, which, however, incorporates essential 
features of hydrodynamical calculations. Let us briefly recall the main 
elements. 

Equilibration of the system is assumed at a formation time 
$\tau_0$, after which isentropic expansion proceeds 
at fixed entropy per baryon, $S/N_B=s/n_B$, where 
the total entropy $S=s V_{FB}$ and net baryon number $N_B=n_B V_{FB}$
are related to the pertinent densities via the time-dependent 
3-volume $V_{FB}$. The latter is modeled in cylindrical symmetry as  
\begin{equation}
V_{FB}(\tau) = 2 (z_0 + v_z \tau +\frac{1}{2}a_z \tau^2)
\pi(r_0+\frac{1}{2}a_{\perp}\tau^2)^2
\label{eq:vol}
\end{equation}
(the overall prefactor of 2 accounts for two fireballs which enable
to cover about 4 units of rapidity).  
$r_0$ denotes the initial transverse overlap of the two colliding nuclei
at given impact parameter $b$, whereas the expansion parameters 
$\{v_z, a_z, a_{\perp}\}$ are adjusted in line with hydro-calculations 
to reproduce observed flow velocities in
connection with (thermal) freezeout times of $\tau_{fo}\simeq 10-14$~fm/c. 
The parameter $z_0$ is equivalent to the formation time $\tau_0$ (in the  
Bjorken limit $z_0 \simeq \tau_0 \Delta y$), specifying the initial 
conditions of the evolution. 
The temperature of the expanding matter is inferred from the   
entropy density $s(\tau)=S/V_{FB}(\tau)$ in either hadronic
or QGP phase (calculated from an ideal resonance gas for the former
and using massive quasiparticles according to Eqs.~(\ref{eq:thmass}) for 
the latter). If $s(\tau)$ lies in between the entropy densities 
$s_c^{H}$ and $s_c^{QG}$ for hadron gas and QGP, a standard mixed phase 
construction~\cite{McLe86} is employed at the critical temperature $T_c$,    
\begin{equation}
\label{eq:hadfrac}
\frac{S}{V_{FB}(t)} = f s_c^{H} + (1-f) s_c^{QG} \ ,  
\end{equation}
which determines the volume partitions  $f$ and $1-f$ for hadronic 
and quark-gluon matter, respectively.  

The collision-energy dependence of the underlying parameters from SPS 
to RHIC has been fixed as follows.  
The total entropy at given collision energy and centrality is obtained
from the specific entropy $S/N_B$ ranging from 26 to 250 from SPS 
to RHIC according to standard chemical freezeout 
analyses~\cite{pbm96-99,pbm01} (with effectively conserved numbers
of pions, kaons etc., thereafter). The critical temperature 
is assumed to increase smoothly from 170~MeV at SPS to 180~MeV
at RHIC (with $\mu_B$ decreasing from 260 to 27~MeV). In addition, we 
account for 10-20~\% variations in $S/N_B$ with centrality following   
particle production systematics reported by NA49~\cite{na49qm99} and 
PHOBOS~\cite{phobos02}. Finally, the formation time $\tau_0$ (\ie, $z_0$)
is continuously decreased from 1 to 1/3~fm/c for 
$\sqrt{s_{NN}}$ from 17 to 200~GeV. 
With the volume expansion as given by Eq.~(\ref{eq:vol}), this results in 
QGP phase durations in central collisions between 1 and  
$\sim$~3~fm/c followed by mixed phases lasting 3-4~fm/c. 

We are now in position to calculate the desired survival
probabilities. The evolution
equation for the number of each charmonium species $i$ ($i=J/\Psi, \Psi',
\chi)$ in the system at time $\tau$ reads
\begin{equation}
  \label{eq:evol}
  \frac{dN_i(\tau)}{d\tau} = - \Gamma_{diss}^i N_i(\tau) \ .
\end{equation}
The destruction rate $\Gamma_{diss}^i$ 
is specified according to the phase and temperature $T(\tau)$ of the
system,  
\begin{equation}
  \label{eq:gamma}
  \Gamma_{diss}^i = \left\{\begin{array}{l l}
      \Gamma_{QG}^i \ , &  T>T_c \\
      f\Gamma_{H}^i + (1-f)\Gamma_{QG}^i  \ , \qquad &  T=T_c \\
      \Gamma_{H}^i \ , &  T<T_c
      \end{array}\right.
\end{equation}
(with $f$ given by eq.~(\ref{eq:hadfrac}), and $\Gamma_{QG}^i$ and
$\Gamma_{H}^i$ from Sects. \ref{sec:qgpsupp} and \ref{sec:hadsupp}). 
Eq.~(\ref{eq:evol}) is readily integrated to obtain the survival 
probability at time $\tau$ during the collision,  
\begin{equation}
\label{eq:survival}
\mathcal{S}_{QG+H}^i(\tau) = e^{-\int\limits_0^\tau 
           \Gamma_{diss}^i(\tau')d\tau'} \ .
\end{equation}
The final survival probability, $\mathcal{S}_{QG+H}$, relevant for experimental 
observables is simply given by the value of $ \mathcal{S}_{QG+H}^i(\tau)$
at the moment of thermal freezeout, $\tau_{fo}$, where all hadrons cease
to interact. Note that $\mathcal{S}_{QG+H}^i$ depends on the impact parameter
$b$ of the collision through the space-time evolution of the system. 

\begin{figure}[htb]
\includegraphics[width=0.45\textwidth,clip=]{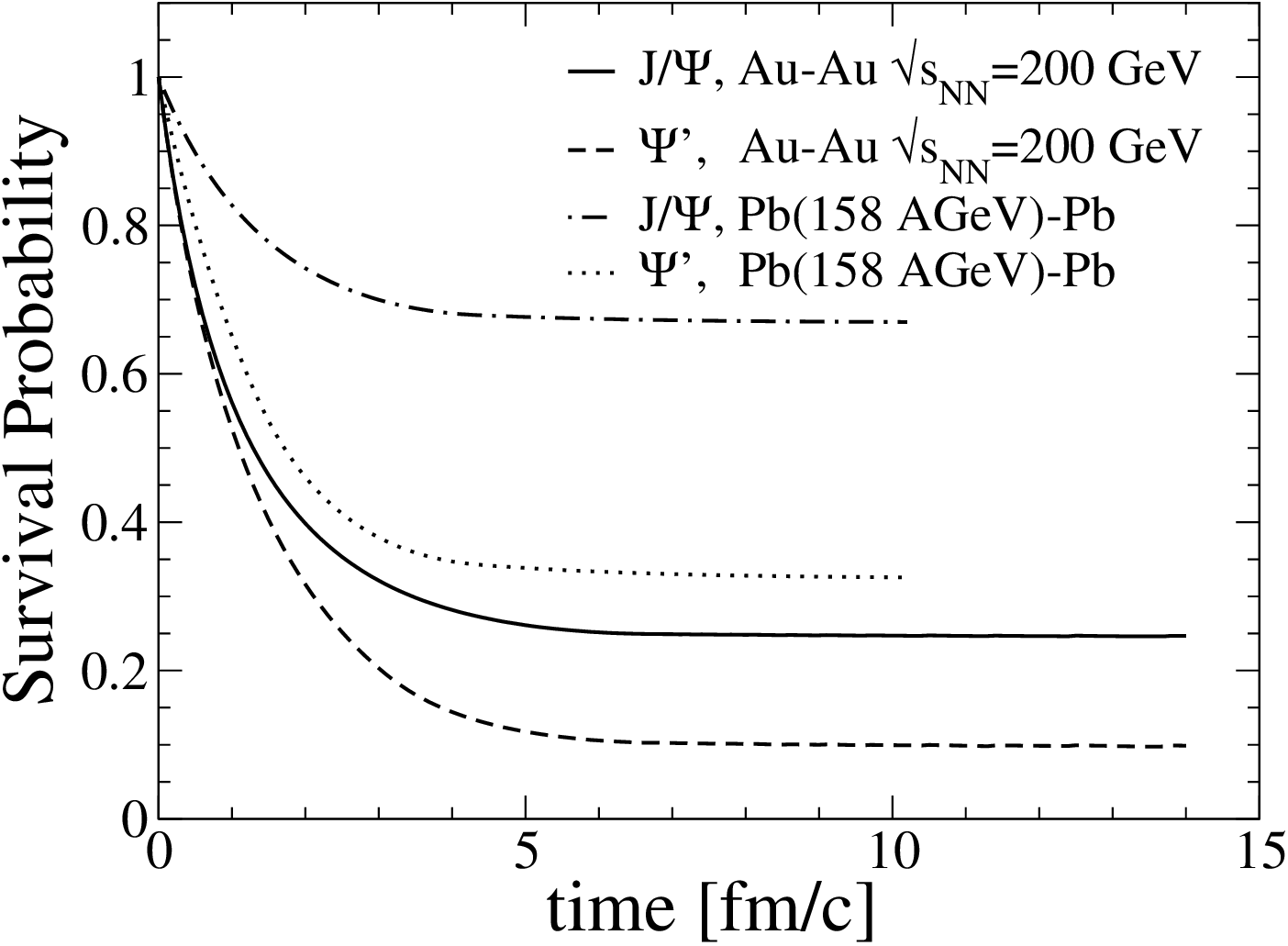} 
\caption{Survival probability of \jj\ (dot-dashed line: SPS, solid
  line: RHIC) and $\Psi'$ (dotted line: SPS, dashed line: RHIC) for
  central collisions ($N_{part}=360$) as a function of time. The
  curves are obtained upon integration of the dissociation rates 
calculated in Sect.~\ref{sec:qgpsupp} and \ref{sec:hadsupp}.}
\label{fig:supp_vs_time}
\end{figure}

Fig.~\ref{fig:supp_vs_time} shows \jj\ and $\Psi'$ survival probabilities 
at full SPS and RHIC energies (central collisions, $N_{part}=360$)
as a function of time with a value of $g=1.7$ for the strong coupling 
constant (which provides reasonable agreement with SPS data to be 
discussed below).
Most of the suppression originates from the plasma and mixed phase, with 
hadronic effects playing little role. 
The destruction of $\Psi'$ states (dotted line) under SPS conditions is 
far from complete (with a fraction of about $1/3$ remaining at the end of 
the collision); this may in fact point at a lack of our understanding
of the hadronic $\Psi'$ interactions, since (i) it will cause problems
in describing the measured $\Psi'/\Psi$ ratio (see Sect.~\ref{sec:psip}
below), and (ii) it appears to be incompatible with recent lattice 
calculations~\cite{pds01} which exhibit a dissolution of the $\Psi'$ in 
the hadronic phase (although no statement about timescales could be made).  
As expected, the overall dissociation is much stronger at RHIC 
(full line: \jj\, dashed line: $\Psi'$), although $25\%$ of 
primordial \jj 's endure the evolution. 

For a combination of the (suppressed) primordial production with the 
statistical one (to be discussed in the next section) it is 
necessary to convert the survival probabilities into absolute yields. 
At a given impact parameter $b$, the number of \jj\ mesons initially 
produced in hard parton-parton collisions is 
\begin{equation}
  N_{J/\Psi}^0(b) = \sigma_{pp}^{J/\Psi} A  B T_{AB}(b) \ .
\end{equation}
There are no data on charmonium production in $p$-$p$ in the RHIC energy  
regime (to date, the highest $cms$ energies available  
lie around $\sqrt{s_{NN}} \simeq 40$~GeV). Thus we have to rely on 
extrapolations. 
The so-called Schuler parameterization~\cite{L96} -- a phenomenological 
fit based on low-energy systematics -- leads to
$\sigma_{pp}^{J/\Psi}(\sqrt{s}=200\;\mbox{GeV})=1.0~\mu$b. On the other
hand,  the simple ansatz $\sigma_{pp}^{J/\Psi}=f\sigma_{pp}^{c\bar{c}}$ 
with $f \simeq 0.025$ \cite{Gav95} together with PYTHIA extrapolations 
of open-charm production (see also next section), which also reproduces 
low-energy data, gives 
$\sigma_{pp}^{J/\Psi}(\sqrt{s}=200\;\mbox{GeV})=8.75~\mu$b. 
This obviously introduces appreciable uncertainties in establishing the
\jj\ yield arising from primordial production mechanisms.

In addition, higher charmonium states 
(in particular $\Psi'$ and various $\chi$ states)  
contribute significantly to the measured \jj\ abundance 
via strong and electromagnetic decays (feeddown). 
Since these resonances have different hadronic and
plasma cross sections, the distinction between prompt and secondary
\jj's has to be made. Consequently, the total number of \jj's 
stemming from primordial charmonium states after
nuclear, plasma and hadronic suppression, becomes  
\begin{equation}\label{eq:totaldirect}
  N_{J/\Psi}^{dir}(b) =
  \sigma_{pp}^{J/\Psi} A B T_{AB}(b)
  \mathcal{S}_{nuc} \left[0.6 \mathcal{S}_{QG+H}^{J/\Psi} + 0.08
    \mathcal{S}_{QG+H}^{\Psi'} + 0.32 \mathcal{S}_{QG+H}^{\chi}\right] \ . 
\end{equation}

\section{Statistical production of $J/\Psi$}
\label{sec:thermal}
Besides direct \jj\ production, another source of \jj's has been attributed  
to thermal (or statistical) production in several variants. We here adopt 
the approach put forward in Refs.~\cite{BS00a,GKSG01}, where  
statistical coalescence of primordial  $c$ and $\bar{c}$ quarks
is assumed to predominantly occur at the hadronization transition. The  
relative abundances of charmed particles are then determined by hadronic 
thermal weights at the critical temperature $T_c$, whereas
the absolute number is specified by the amount of charm quarks 
created in the early (hard) $N$-$N$ collisions.  

Starting from the ideal gas expression for the thermal number 
density of particle species $i$,   
\begin{equation}
  n_i = \frac{d_i}{2\pi^2}\int\limits_{0}^{\infty}
  p^2dp\left[\exp\left(\frac{\sqrt{p^2+m_i^2}-\mu_i}{T}\right)\pm1\right]^{-1} 
\end{equation}
($d_i$: spin-isospin degeneracy), 
the total number densities of open and hidden charm particles follow as   
\begin{eqnarray}
n_{op} &=& \sum\limits_{}^{} n_i \ , \qquad i=D,D^{\star},\cdots
\label{eq:nop}
\\
n_{hid} &=& \sum\limits_{}^{} n_j \ , \qquad j=\eta_c,\Psi,\cdots \ . 
\label{eq:nhid}
\end{eqnarray}
The pertinent chemical potentials $\mu_i$ encode conserved 
charges ({\em net} baryon, strangeness and charm numbers), and depend
on the collision energy of the system (in accord with the hadro-chemistry).  
At $T_c$ we match the densities, Eqs.~(\ref{eq:nop}) and (\ref{eq:nhid}), 
to the available number
of open charm pairs, $N_{c\bar c}$, by introducing an effective fugacity
factor $\gamma_c = \gamma_{\bar{c}}$ for both charm and anticharm 
quarks. Since $N_{c\bar c}$ is typically a small number (\eg, 
$\sim$0.2 for central $Pb$-$Pb$ at SPS), exact charm conservation within
the canonical ensemble formalism~\cite{Shu75,RT80} is mandatory, 
leading to     
\begin{equation}
\label{mastereq}
N_{c\bar{c}} = \frac{1}{2}\gamma_c n_{op} V_{H}
\frac{I_1(\gamma_c n_{op} V_{H})}{I_0(\gamma_c n_{op} V_{H})}
+ \gamma_c^2 n_{hid} V_{H} \ .  
\end{equation}
Here, $V_H$ denotes the fireball volume from the previous section, 
cf.~Eq.~(\ref{eq:vol}), at the moment when hadronization is complete. 
The actual value of $N_{c\bar{c}}$ depends on both collision energy 
and impact parameter. The former dependence is inferred from  
PYTHIA computations~\cite{pythia} upscaled by an empirical 
$K$-factor of around 5 extracted from a best fit to existing 
$pN$ and $\pi N$ data~\cite{AD00}. As discussed in the previous 
section, the extrapolation into the RHIC regime bears appreciable 
uncertainty. {\em E.g.}, next-to-leading order pQCD
calculations~\cite{McGAU95} for full RHIC energy give 
$\sigma_{pp}^{c\bar{c}} \sim 350$ $\mu$b, to be compared to 
$\sigma_{pp}^{c\bar{c}} \sim 570$ $\mu$b from the PYTHIA estimate. 
Both extrapolations are in line with recent indirect measurements 
from PHENIX~\cite{Phenix02} in $Au$-$Au$ at $\sqrt{s_{NN}}=130$~GeV, 
where single-electron $p_t$-spectra have been used to infer  
$\sigma_{pp}^{c\bar{c}} =380\pm60(\mbox{stat})\pm200(\mbox{syst})$ $\mu$b,
to be compared to $\sigma_{pp}^{c\bar{c}}\sim 320$ $\mu$b within the 
PYTHIA extrapolation. 

By construction, statistical charmonium production 
is only active for  
$c$ and $\bar c$ that emerge from a deconfined environment prior to 
recombination. Therefore, in peripheral collisions where the initial 
volume is only partially filled with the QGP phase, only a fraction of
the $c\bar{c}$ pairs is available for coalescence. Accordingly, the
hadronization volume which enters Eq.~(\ref{mastereq}) arises from  
the initial volume in the plasma phase after expansion.  

The thermal equilibrium (but chemical off-equilibrium)  \jj\ yield 
takes the form  
$\langle J/\Psi \rangle = \gamma_c^2~V_{H}~n_{J/\Psi}$.    
The former expression is valid if 
(anti-) charm quarks are kinetically  
equilibrated, \ie, the momentum distribution of $c$ and $\bar{c}$
quarks is thermal which is questionable under SPS and even
RHIC conditions.  We therefore implement the following correction: 
we introduce a thermalization time $\tau_{eq}$ for $c$ and $\bar{c}$ 
quarks, approximated as $\tau_{eq} = 1/n\sigma$, where $n$ is the total 
density of quark- and gluon-quasiparticles in the system, and
$\sigma$ is the elastic scattering cross section for the processes
$g(q,\bar{q}) + c(\bar{c}) \rightarrow g(q,\bar{q}) + c(\bar{c})$. 
Within a relaxation time approach, the relative reduction 
$\mathcal{R}$ in  thermal \jj\ formation is then estimated as 
\begin{equation}
  \mathcal{R} = \left[1-\exp\left(-\int\limits_{0}^{\tau_{H}}
      \frac{d\tau}{\tau_{eq}}\right)\right] \ ,   
\end{equation}
where $\tau_{H}$ is the time at which hadronization is completed. 
Our expression for the 
number of \jj's produced at the hadronization transition by
coalescence of a $c$ and $\bar{c}$ quark is thus modified according to 
\begin{equation}
\langle J/\Psi \rangle = \gamma_c^2\; V_{H}\; n_{J/\Psi} \mathcal{R} \ . 
\end{equation}
We have checked that our results do not change significantly
with the definition adopted for $\tau_{H}$. The latter is not  
a sharply defined quantity since the mixed phase lasts for a few
fm/c. However, smaller values of $\tau_{H}$ (\eg,  if 
taken in the middle of the mixed phase) lead to smaller volumes and
larger $\gamma_c$ implying an increase in thermal \jj\
production which, in turn, is (partially) compensated by a less degree of 
thermalization through a smaller value of $\mathcal{R}$. 

The charmonia statistically produced at the hadronization transition are
still subject to reinteractions in the hadronic phase, so that their final
contribution to the observed \jj\ yield is accounted for via  
\begin{equation}\label{eq:totalthermal}
  N_{J/\Psi}^{th} = \gamma_c^2 \ V_{H}
  \left[n_{J/\Psi} \ \mathcal{S}_{H}^{J/\Psi} + \sum\limits_{X}^{}
    BR(X\rightarrow J/\Psi) \ n_X \  \mathcal{S}_{H}^{X}\right] 
 \ \mathcal{R} \ , 
\end{equation}
where the summation is carried over the charmonium states X with finite
decay branching into \jj's (feeddown).

\section{Two-component model at the SPS}
\label{sec:sps}
By combining the two sources of charmonium as discussed in the previous 
sections, we now turn to applications to heavy-ion collisions
at SPS energy ($\sqrt{s}_{NN}=17.2$~GeV).  
We first address the observable that has drawn the most attention, \ie, 
the centrality dependence of the \jj\ yield, but also investigate 
the $\Psi'/\Psi$ ratio. 

\subsection{Centrality dependence of \jj\ production}
One of the important findings of the NA38 and NA50 experiments at CERN is 
that the \jj\ yield in $p$-$p$ and $p$-$A$ collisions with light and heavy 
targets is well explained by hard production coupled with nuclear 
absorption, cf. Sect.~\ref{sec:nucsupp}. Thus, any
attempt to describe the \jj\ yield in heavy-ion collisions must
reproduce this feature. In particular, for very peripheral collisions 
involving only a few participant nucleons, the \jj\ suppression pattern 
should coincide with nuclear absorption. 

In our two-component model, the total number of observed \jj's
is the sum of direct and statistical production according to
Eqs.~(\ref{eq:totaldirect}) and (\ref{eq:totalthermal}), respectively,  
\begin{equation}
N_{J/\Psi} = N_{J/\Psi}^{dir} + N_{J/\Psi}^{th} \ .
\end{equation}
Both terms on the right-hand side depend on centrality (via the 
impact parameter $b$) and collision-energy. For very peripheral 
collisions in the SPS regime, 
the initial conditions are not energetic enough to induce a transition
into the QGP. Therefore, the \jj\ source of statistical recombination 
at $T_c$ is absent; at the same time the hadronic interactions are not
frequent enough to induce sizable effects. Thus, for large impact
parameters our approach is consistent with the NA38/NA50 results.

\begin{figure}[!h]
\centering
\mbox{\subfigure{\epsfig{figure=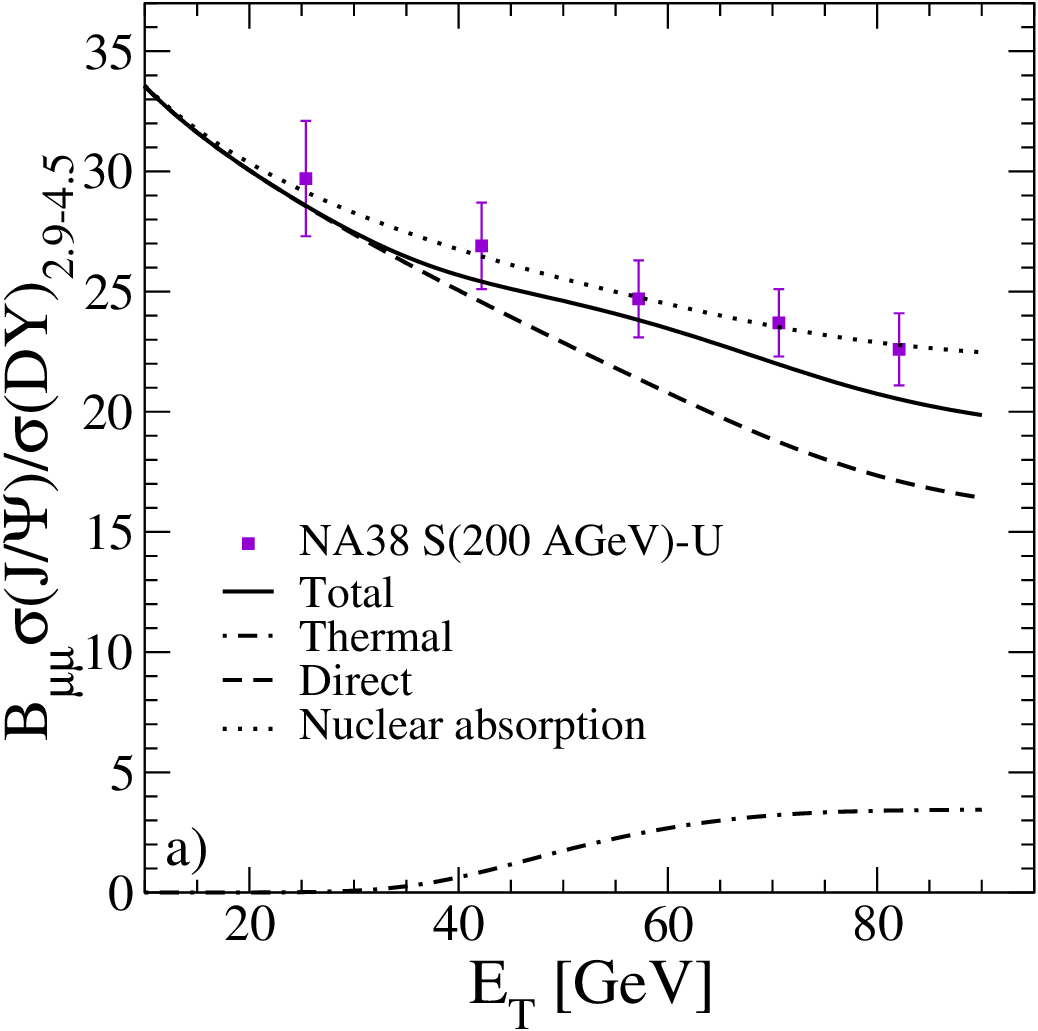,width=0.385\textwidth,clip=}}
  \quad \subfigure{\epsfig{figure=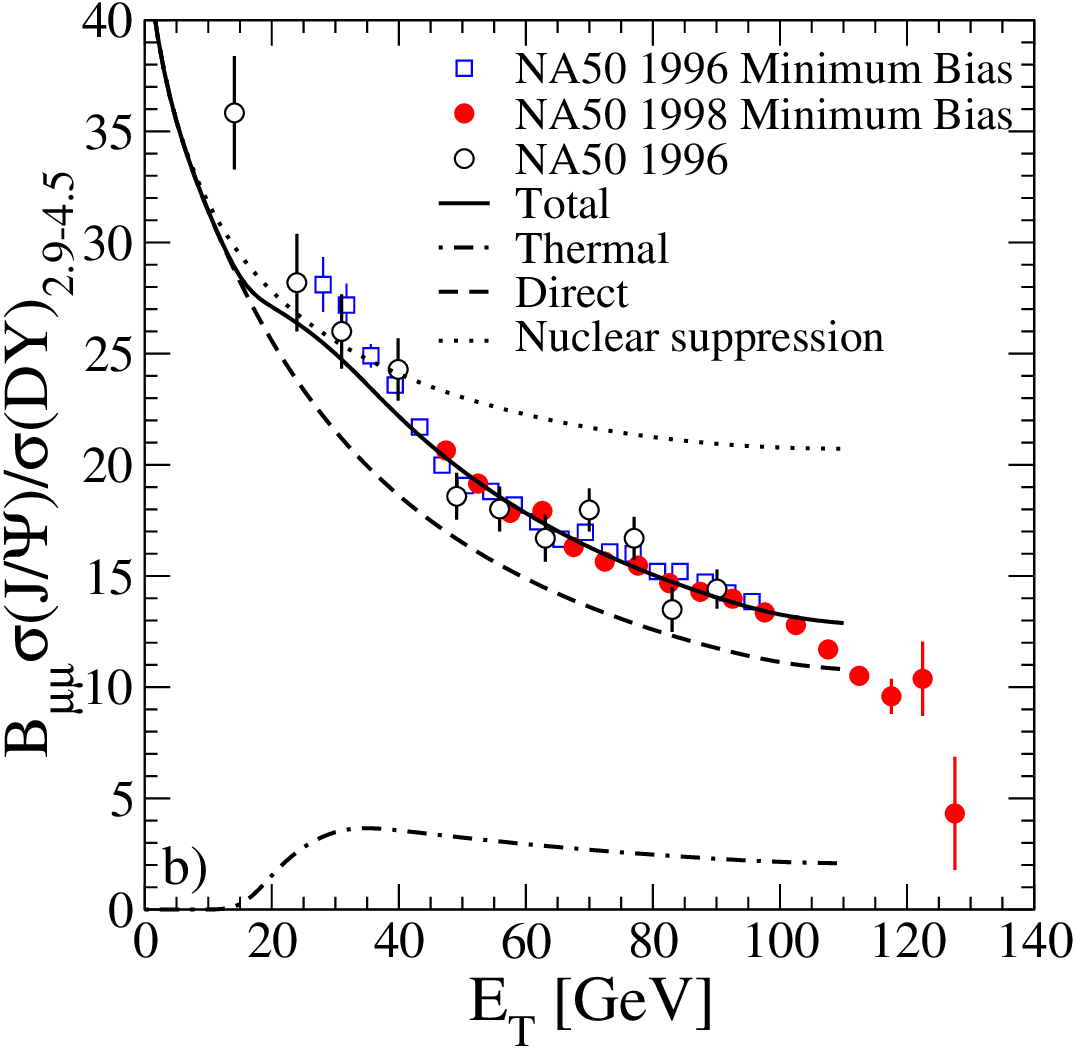,width=0.4\textwidth,clip=}}}
\caption{Results on \jj\ production within the two-component model as a 
function of centrality compared to NA38~\cite{na38-99a} (left panel)
and NA50 data~\cite{na50-00a} for $S$(200 AGeV)-$U$ and $Pb$(158 AGeV)-$Pb$,
respectively. 
Dotted lines: direct production with nuclear absorption alone;
dashed lines: direct production subject to nuclear, QGP and hadronic
absorption; dot-dashed lines: statistical (thermal) production
from a hadronizing QGP including hadronic dissociation;   
solid lines:  combined direct and statistical yield (sum of dashed 
and dot-dashed lines).}  
\label{fig:na38_na50}
\end{figure}

For a detailed comparison with data, we need to 
evaluate the ratio $B_{\mu\mu}\sigma^{J/\Psi}/\sigma^{DY}$ commonly 
displayed by NA38/NA50 as a function of transverse energy deposited 
in their calorimeter.  
Following the treatment outlined in Sect.~\ref{sec:nucsupp}, 
we convolute our theoretical results (which are obtained as a function
of impact parameter $b$) with the probability distribution 
$\mathcal{P}(E_T,b)$. This amounts to replacing the numerator in  
Eq.~(\ref{eq:JDYrat}) by the expression 
\begin{equation}
B_{\mu\mu} \sigma_{J/\Psi}(E_T) = 
B_{\mu\mu}\sigma_{J/\Psi}^{pp}
 \int\limits_{}^{}d^2b \ {\cal P}_{AB}(E_T,b)
\left[ {\cal S}_{nuc} {\cal S}_{QG+H} + 
N_{J/\Psi}^{th} / (\sigma^{pp}_{J/\Psi} AB T_{AB}(b)) \right] T_{AB}(b) \ .  
\end{equation}
Our results are compared to NA38/NA50 data in Fig.~\ref{fig:na38_na50}
for both the $S$(200 AGeV)-$U$ (left panel)  and $Pb$(158 AGeV)-$Pb$
(right panel) system.
At all centralities, direct production (dashed line) prevails over the thermal
component (dot-dashed line). The latter sets in once a QGP starts forming, 
which, in turn, requires a stronger QGP suppression of the direct component 
than without the thermal contribution.
The adjustment of the only free parameter (strong coupling constant  
$g=1.7$) to the most central $Pb$-$Pb$ data allows for a satisfactory 
reproduction of the centrality dependence for this system. 
For $S$-$U$ collisions, the results are somewhat on the low side. 
Note that in our approach the ``drop'' in the $Pb$-$Pb$ data around
$E_T \simeq 40$ GeV is a combination of a rather strong QGP
suppression coupled with the onset of thermal production.
  
In its present form, our model does not capture the appearance of the 
``second drop'' in the data for the most central $Pb$-$Pb$ collisions 
at $E_T>100$~GeV.  In fact, the maximum transverse energy in our
description is at $E_T^{max}=E_T(b=0)=100$~GeV, well below the
experimental limit which extends up to $E_T\simeq 125$~GeV. 
It has been suggested that these features are associated with  
transverse energy fluctuations~\cite{CFK00,BDO00,HKP00} 
and/or trigger energy losses~\cite{CKS02}, and are thus not 
necessarily related to a shortcoming in an underlying (microscopic) 
model description of $J/\Psi$ production (suppression). 

\begin{figure}[!h]
\centering
\mbox{\subfigure{\epsfig{figure=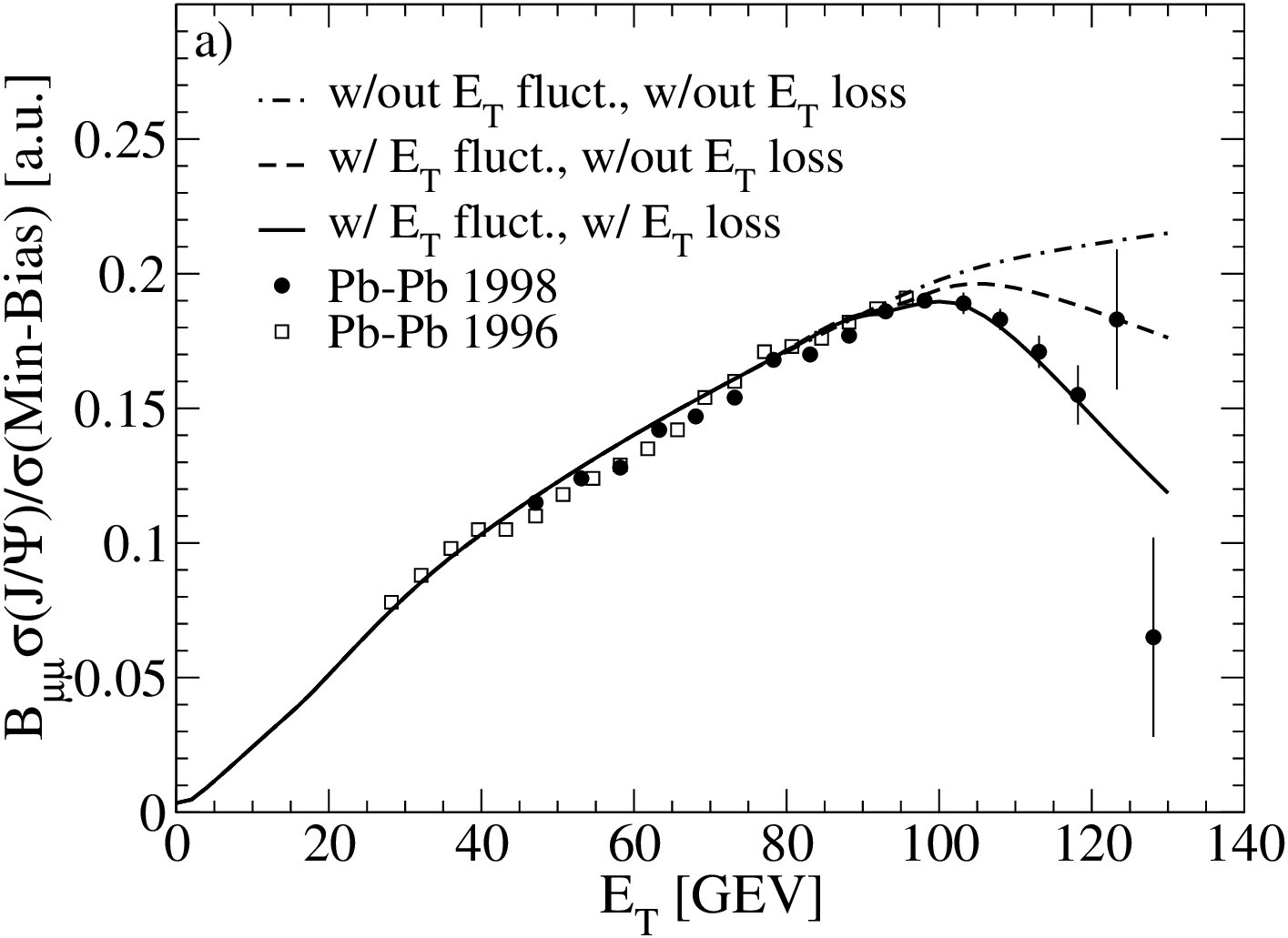,width=0.423\textwidth,clip=}}
  \quad \subfigure{\epsfig{figure=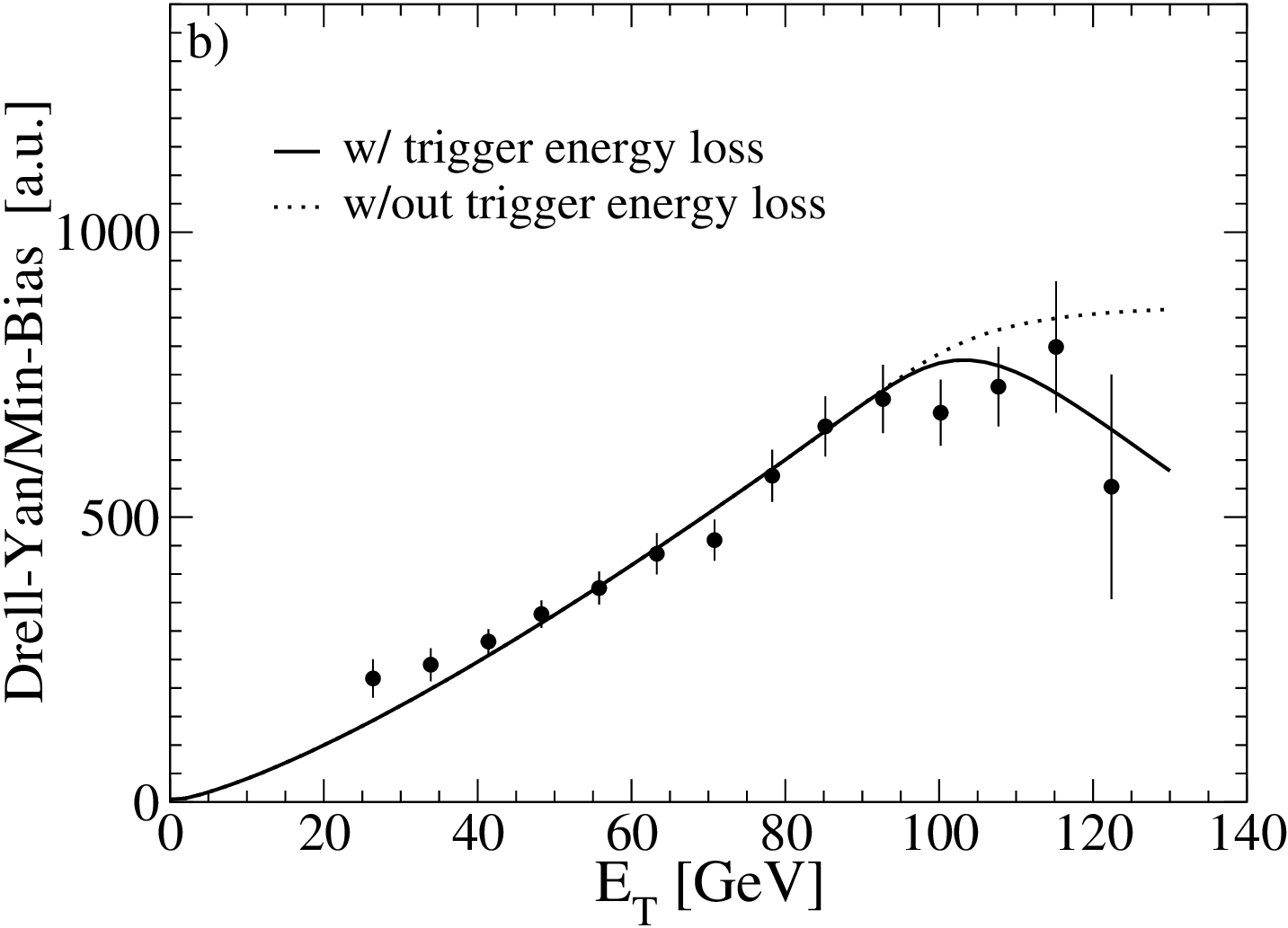,width=0.423\textwidth,clip=}}}
\caption{Left panel: \jj\ over Minimum Bias ($MB$) ratio compared
to calculations within the two-component model (dot-dashed curve) 
with additional inclusion of effects due to $E_T$ fluctuations (dashed curve)
and trigger energy loss (solid curve). Right panel: impact of a model 
for trigger energy loss on the Drell-Yan ($DY$) over $MB$ ratio. 
All data points are from NA50 for $Pb$(158 AGeV)-$Pb$~\cite{na50-99a}.} 
\label{fig:j_mb_dy}
\end{figure}

Let us first address the $E_T$ fluctuations. 
From the minimum bias ($MB$) event distribution of transverse energy, 
$dN/dE_T$, as measured in the NA50 apparatus~\cite{na50-99a} one finds a 
rapid falloff beyond  $E_T=100$ GeV, the so-called ``knee'' of the 
distribution. The tail of the latter, which reaches beyond
$E_T=100$ GeV, is  associated with fluctuations in transverse 
energy at fixed geometry for $b=0$. Events which fluctuate  
beyond $E_T>100$~GeV contain a larger initial
entropy and consequently a hotter initial temperature
and a longer plasma phase which implies additional \jj\ 
suppression (charm and $J/\Psi$ production being a hard process are 
not coupled to fluctuations in the ``soft'' sector). 
To account for this phenomenon~\cite{CFK00}, we replace in our calculations 
the total entropy at fixed impact parameter, $S_{tot}(b)$, by
\begin{equation}
  S_{tot}(b) \rightarrow S_{tot}(b)\frac{E_T}{E_T(b)} \ .
\end{equation}
This does not affect our results for $E_T\le E_T(b=0)\simeq 100$~GeV, 
but beyond the knee of the distribution, $S_{tot}$ 
is enhanced by the factor $E_T/E_T(b)$. 
The effect of this modification is shown in the left panel of 
Fig.~\ref{fig:j_mb_dy}, where we compare our calculations to data
on $J/\Psi$ production normalized to the minimum bias 
$E_T$-distribution\footnote{This way of normalizing the data has the 
advantage of the much better statistics for $dN/dE_T$ as compared 
to the Drell-Yan sample. On the other hand, since  $E_T$ production
is governed by soft physics (which essentially scales with the number
of participants rather than the number of $N$-$N$ collisions), 
the characteristic features of $J/\Psi$ suppression are not readily
discernible.}. Obviously, the description of the observed turnover for 
$E_T>100$~GeV is improved by inclusion of $E_T$-fluctuations 
(cf.~dashed versus dot-dashed curve), but does not
suffice to quantitatively  explain the data. Note that this effect 
relies on additional $J/\Psi$ suppression in both the direct 
component (due to stronger suppression) and the thermal 
component (due to a larger hadronization volume weakening 
the enhancement induced by the canonical-ensemble treatment). 
  
However, as has been suggested by Capella \etal~\cite{CKS02}, there
might be an additional feature in the large $E_T$-region unrelated
to $J/\Psi$ physics, which can be gleaned from the Drell-Yan (DY) over
minimum bias data, cf.~right panel of Fig.~\ref{fig:j_mb_dy}.      
The theoretical ratio $DY/MB$ (dotted line) computed by NA50 flattens 
out at large values of $E_T>100$ GeV, whereas the data seem to indicate 
a slight turnover~\cite{na50-99a}.
The argument~\cite{CKS02} to explain this turnover   
(which equally applies to the $J/\Psi$ event sample)
is that for DY-events the hadronic transverse energy deposited in the 
calorimeter is slightly reduced as compared to the corresponding  
$MB$-events due to triggering on the $DY$ (or $J/\Psi$) pair 
(2.9~GeV$<M_{DY}<$4.5~GeV). As elaborated in Ref.~\cite{CKS02},  
a rough estimate of this effect is obtained by rescaling the amount of 
transverse energy in the $J/\Psi$ and $DY$ events according to  
\begin{equation}
  E_T(b) \rightarrow E_T(b)\frac{N_{part}-2}{N_{part}} \ . 
\end{equation} 
When incorporated into the $E_T$-$b$ correlation, Eq.~(\ref{eq:PAB}), 
this rather small loss in $E_T$ entails a surprisingly 
large drop of about $\sim$20-30\% in the tail of the $(DY/MB)$ 
distribution, cf.~ Fig.~\ref{fig:j_mb_dy} (solid curve in the right panel).

\begin{figure}[!thb]
\includegraphics[width=0.4\textwidth,clip=]{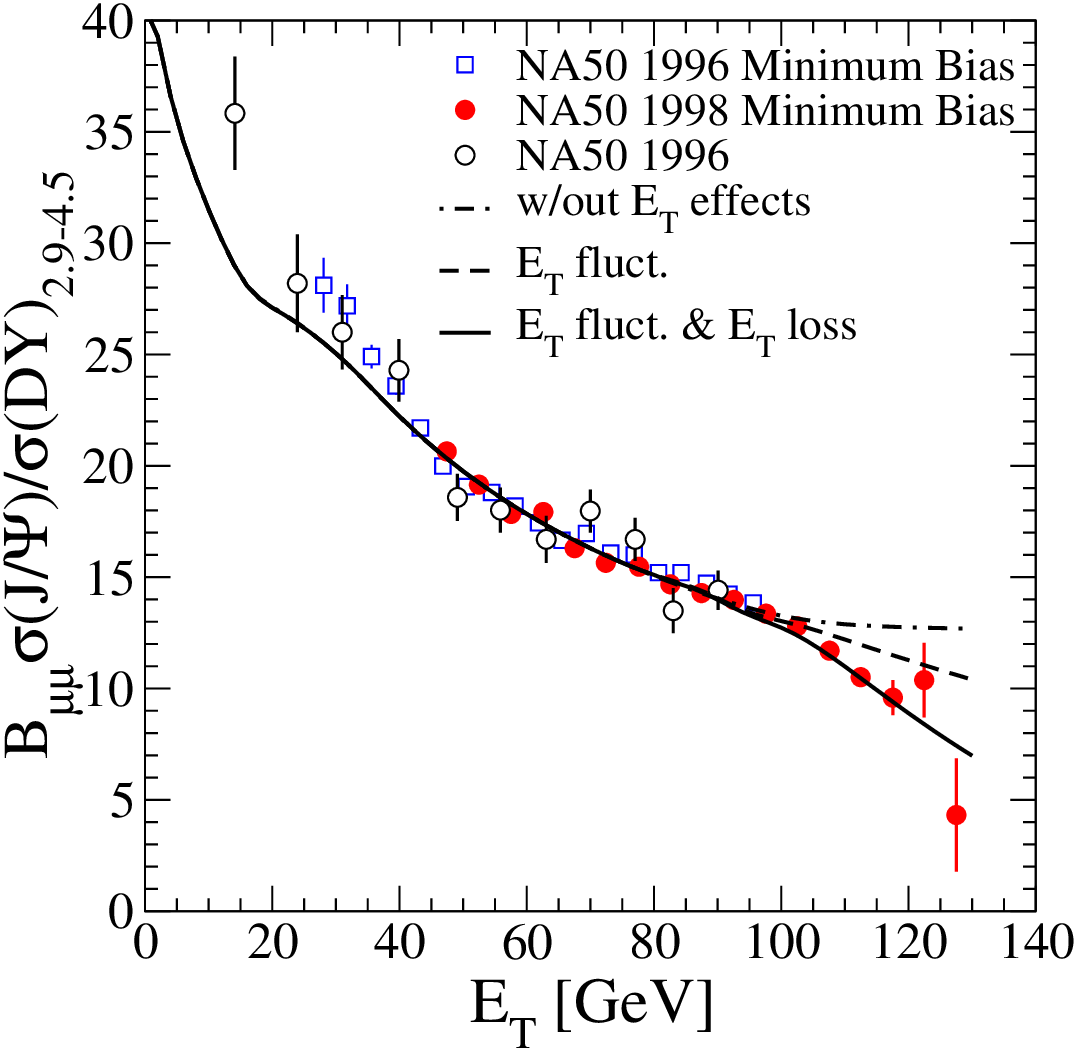} 
\caption{Results of the two-component model without (dot-dashed line) 
and with additional inclusion of transverse
energy fluctuations (dashed line) and trigger energy loss (full
line), for the centrality dependence of the
  $B_{\mu\mu}\sigma^{J/\Psi}/\sigma^{DY}$ ratio in $Pb$(158 AGeV)-$Pb$
  collisions.} 
\label{fig:j_dy_fluc}
\end{figure}

Returning to the $(J/\Psi)/MB$ ratio (left panel of Fig.~\ref{fig:j_mb_dy}), 
we see that the combined effect of $E_T$ fluctuations and trigger
energy loss, implemented within our microscopic model, gives 
a satisfactory description of the experimental findings. This also
holds true for the more common representation of the data 
via the $J/\Psi/DY$ ratio, cf.~Fig.~\ref{fig:j_dy_fluc}. Note that in 
this case the trigger energy-loss
correction only applies to the ``$MB$'' data sets, which have been 
extracted using a theoretical expression for the $MB/DY$ ratio
according to  
\begin{equation}
  \left(\frac{J/\Psi}{DY}\right)_{MB\;analysis} =
  \left(\frac{J/\Psi}{MB}\right)_{exp}\left(\frac{MB}{DY}\right)_{th} \ .
\end{equation}

\subsection{$\Psi'/\Psi$ ratio}
\label{sec:psip}
In $p$-$p$ collisions in the SPS energy regime, the ratio of 
produced $\Psi'$ to $J/\Psi$ mesons amounts to a value of about 
15\%, which persists for $p$-$A$ collisions as nuclear absorption
affects both charmonium states in practically the same way, 
cf.~Sect.~\ref{sec:nucsupp}. A marked deviation from this behavior
has been observed in $S$(200~AGeV)-$U$
and $Pb$(158~AGeV)-$Pb$ collisions, with an onset at rather low
centralities.   Remarkably, for central collisions, the 
$\Psi'/\Psi$ ratio does not go to zero, but rather levels off at 
a value of around 3-4\%.  
This is contrary to the naive expectation that $\Psi'$ mesons, due to 
their much smaller  binding energy than the \jj\ states, are
significantly more suppressed. In Ref.~\cite{SSZ97} it has been 
suggested that, under the premise that most $\Psi'$ states are dissolved
in the QGP, their abundance is regenerated in the hadronic phase
from remaining $\Psi$ states as a consequence of chiral symmetry 
restoration, via the process $\Psi + \pi \rightarrow \Psi'$.  
The interaction was assumed to be mediated through $\sigma(500)$ meson 
exchange, the mass of the latter approaching the pion mass thereby  
substantially enhancing $\Psi'$ formation. From another 
point of view, the fact that the value of 4\%
reflects the thermal ratio of $\Psi'/\Psi$ at a temperature of 
$T=170$~MeV (with little latitude), has been put forward in 
ref.~\cite{BS00a} as evidence for statistical charmonium production
at the hadronization transition. 

\begin{figure}[h!]
\includegraphics[width=0.5\textwidth,clip=]{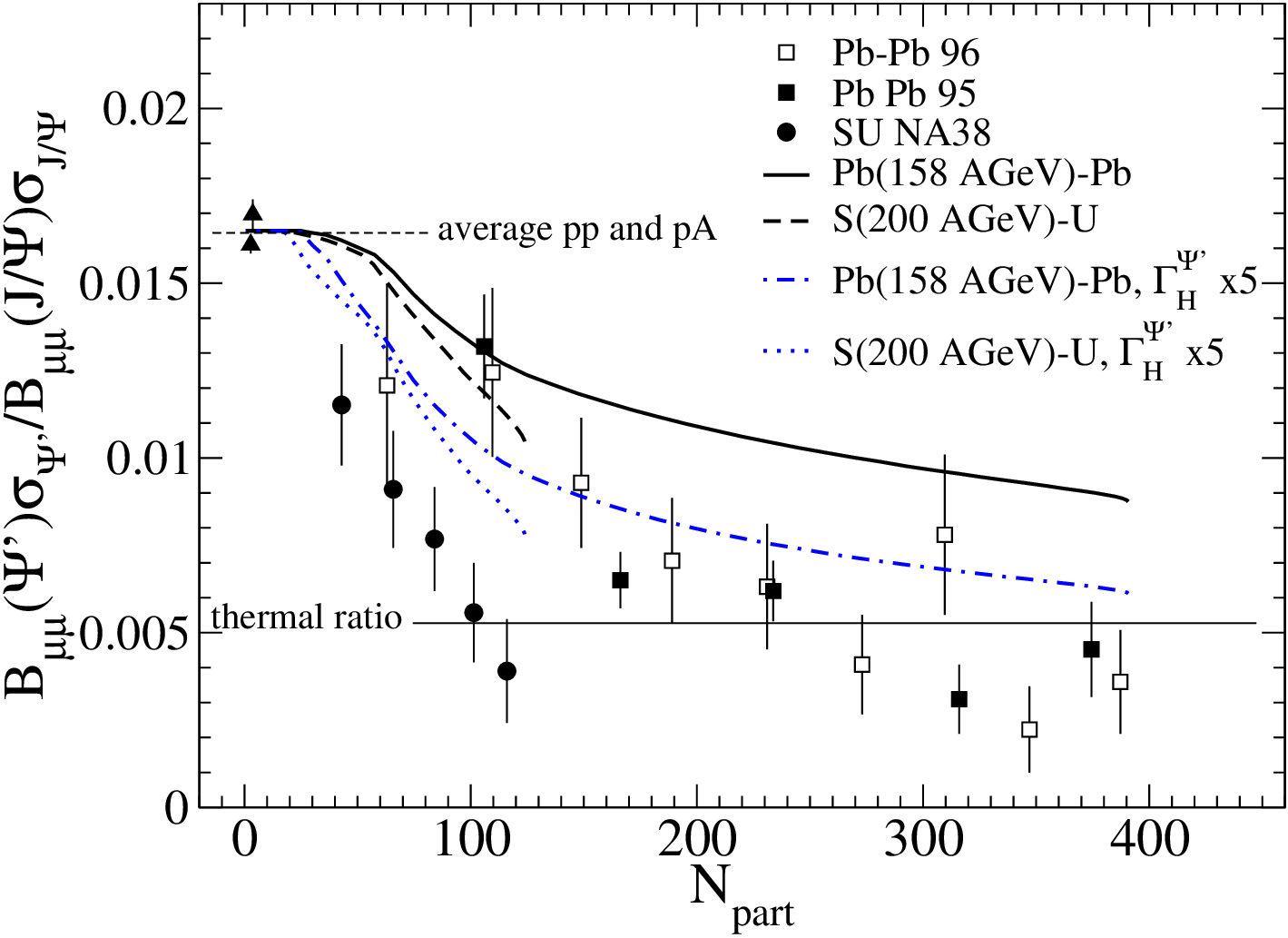} 
\caption{Our calculations for the centrality dependence of $\Psi'/\Psi$
compared to data from NA38/50. The solid and dashed line are for 
$S$(200~AGeV)-$U$ and $Pb$(158~AGeV)-$Pb$, respectively, using hadronic 
$\Psi'$ dissociation cross sections obtained by geometric scaling
of the $J/\Psi$ one, cf.~Fig.~\ref{fig:tauhad}. The dashed and dotted 
lines are the corresponding results when artificially increasing the 
$\Psi'$ cross sections by another factor of 5.}
\label{fig:psip}
\end{figure}

Within our two-component model as laid out above, the $\Psi'/\Psi$ ratio
follows without further assumptions.  
The results for both $S$-$U$ and $Pb$-$Pb$ systems are compared 
to NA38/50 data in Fig.~\ref{fig:psip}. 
The discrepancy with experiment is rather significant, especially 
for $S$(200~AGeV)-$U$. Since in the latter reaction QGP effects are
not expected to play a pronounced role, it seems that the deficiency 
in our description has to be assigned to the hadronic phase, \ie, an 
underestimation of the hadronic cross sections for the $\Psi'$.
Indeed, an artificial increase of this quantity
by, say, a factor of 5 clearly improves the agreement with the data. 
We have checked that such an increase
in the $\Psi'$ hadronic cross sections has negligible impact on the
$\Psi/DY$ ratio as plotted in Fig.~\ref{fig:j_dy_fluc} 
(the $\Psi'$ contributes maximally  $8\%$ to the observed \jj\ yield). 

As mentioned before, recent lattice calculations indicate a dissolution
of the $\Psi'$ in a static environment well below the phase transition
temperature, due to the lowering of the $D \bar D$ continuum threshold
below the in-medium $\Psi'$ mass. In a hadronic model framework, this 
can be implemented by an in-medium reduction of the $D$-meson masses,
which has been motivated in Ref.~\cite{Sib00} by chiral restoration
arguments inducing a lowering of the light-quark mass within the 
$c\bar q$ and $\bar cq$ states. We have investigated this possibility
within the hadronic approach employed here, and found a strong 
sensitivity to the detailed modeling of the light-quark related 
portion of the $D$-meson masses. In fact, if the lifetimes of the
charmonium states become comparable to duration of the fireball   
expansion, one needs to account for the reverse reaction of 
charmonium formation (as required by detailed balance), which is 
beyond the scope of
this paper\footnote{Note that with an increase by a factor of 5 for  
$\Psi'$ dissociation rate over the results shown in 
Fig.~\ref{fig:tauhad} (as applied in Fig.~\ref{fig:psip}) 
the $\Psi'$ lifetimes in the vicinity of $T_c$ are indeed close   
to the expansion time of the hadronic phase.}. 

A more controlled way to assess hadronic medium effects in charmonium 
dissociation should be provided by constituent quark (exchange-) models 
incorporating both phenomenological confinement potentials as well as 
properties of chiral symmetry breaking~\cite{MBQ95,BSW01,BBK01,BSW02}. 
This will be addressed elsewhere~\cite{GRW02}. 

\section{Excitation function and predictions for RHIC}
\label{sec:rhic}
An essential part of the experimental program at RHIC is again  
on (penetrating) electromagnetic probes. The PHENIX
detector will provide accurate dilepton data via both the 
(forward) muon arms as well as electron identification in the central 
region. The results on charmonium should  allow for stringent
constraints on models. 
At full RHIC energy, standard extrapolations predict an open-charm 
production that is about two orders of magnitude larger than in the
SPS regime, entailing a substantial increase in the statistical 
recombination mechanism for charmonia. At the same time, direct 
(hard) charmonium production, albeit
also enhanced by presumably a similar factor as open charm, ought to be 
more strongly suppressed due to longer and initially hotter QGP phases.

\begin{figure}[h!]
\includegraphics[width=0.5\textwidth,clip=]{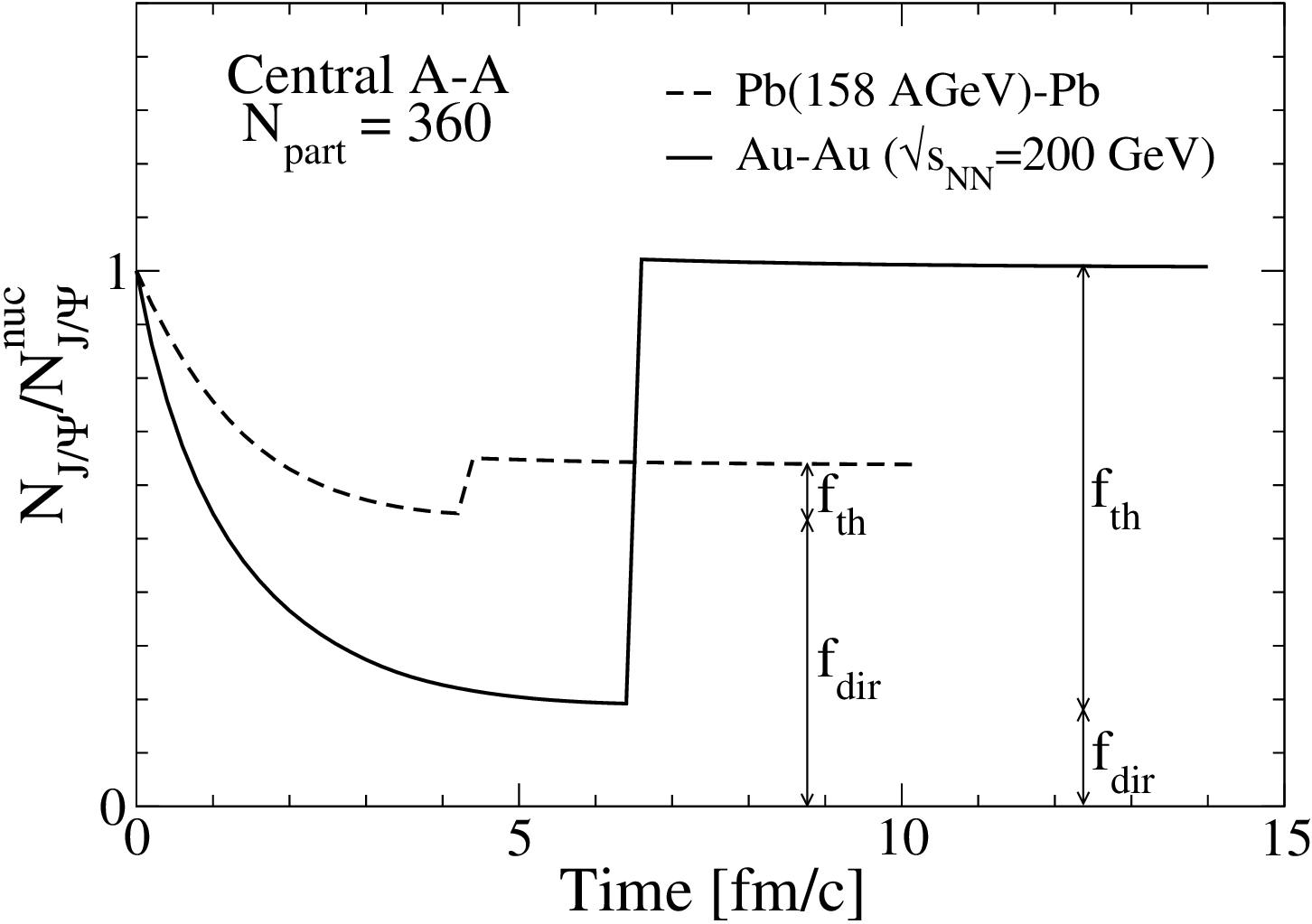} 
\caption{Time dependence of the ratio $N_{J/\Psi}/N_{J/\Psi}^{nuc}$ at
SPS (dashed line) and RHIC (full line) for central collisions with
$N_{part}=360$, where $N_{J/\Psi}^{nuc}$ is the number of \jj's
remaining after nuclear absorption. The respective fractions of
direct ($f_{dir}$) and thermal ($f_{th}$) yields are indicated by 
the arrows.}
\label{fig:Nj_vs_time}
\end{figure}

A quantitative comparison between SPS and RHIC within our two-component
model is performed in Fig.~\ref{fig:Nj_vs_time} where the final (observed) 
number of \jj's, normalized to the number \jj's remaining after 
nuclear absorption, $N_{J/\Psi}^{nuc}$, is displayed for central
collisions as a function of the fireball evolution time. 

The freezeout value for this ratio increases by about 50\%  
going from SPS to RHIC (from 0.65 to about 1). More dramatically,  the 
composition in terms of underlying sources is very different: whereas at 
SPS the (suppressed) direct yield dominates, $J/\Psi$-mesons at RHIC
originate to $\sim$80\% from thermal production (being proportional to 
$(N_{c\bar{c}})^2$). 
The upward jump of the two curves in Fig.~\ref{fig:Nj_vs_time} is 
located at the respective end of the mixed phase, $\tau_H$, where in our 
approximation all thermal production is concentrated. As elucidated  
in Sect.~\ref{sec:thermal}, the final results are not sensitive to the
exact production time within the mixed phase. 

\begin{figure}[!h]
\centering
\mbox{\subfigure{\epsfig{figure=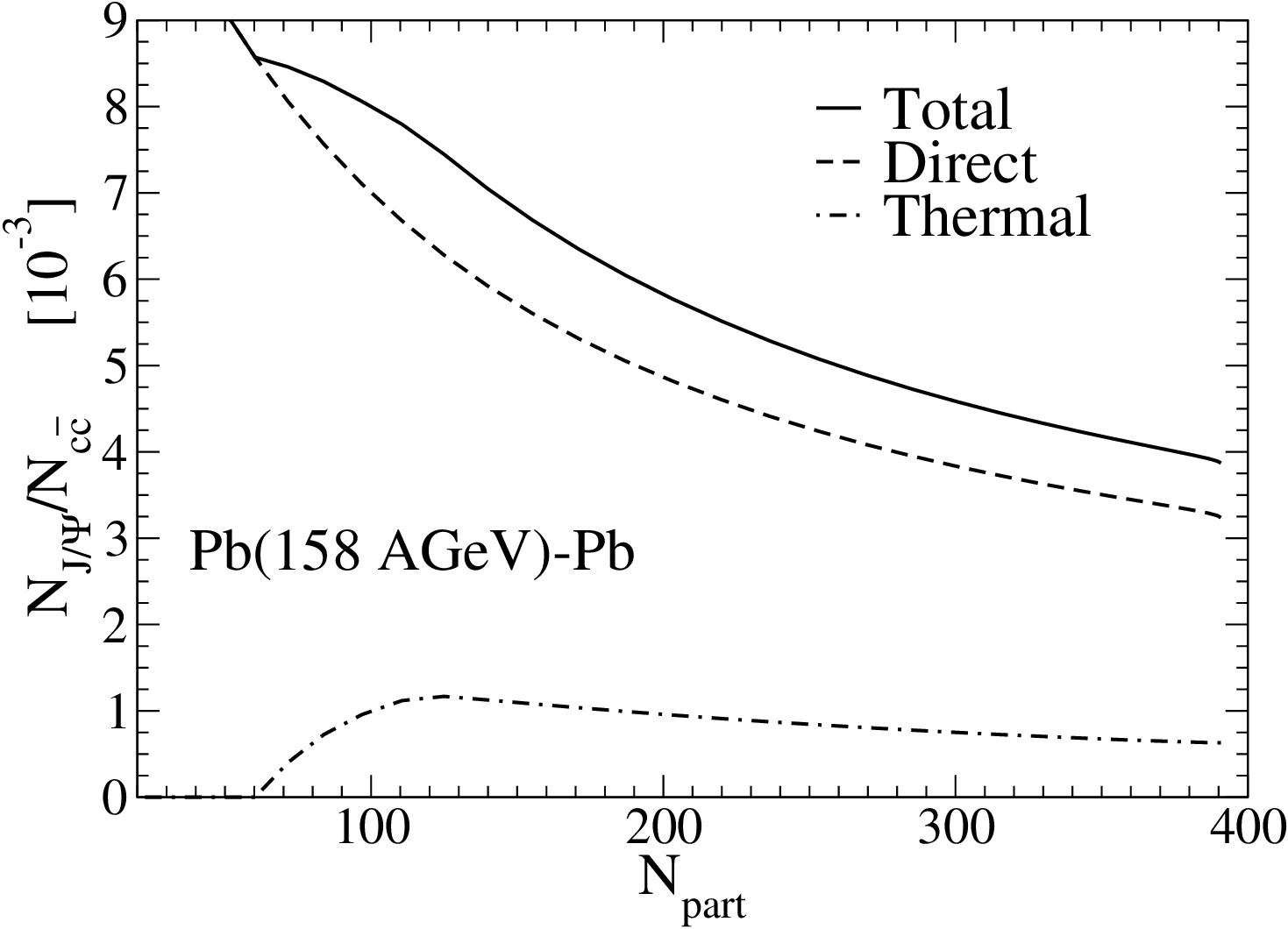,width=0.48\textwidth,clip=}}
  \quad 
\subfigure{\epsfig{figure=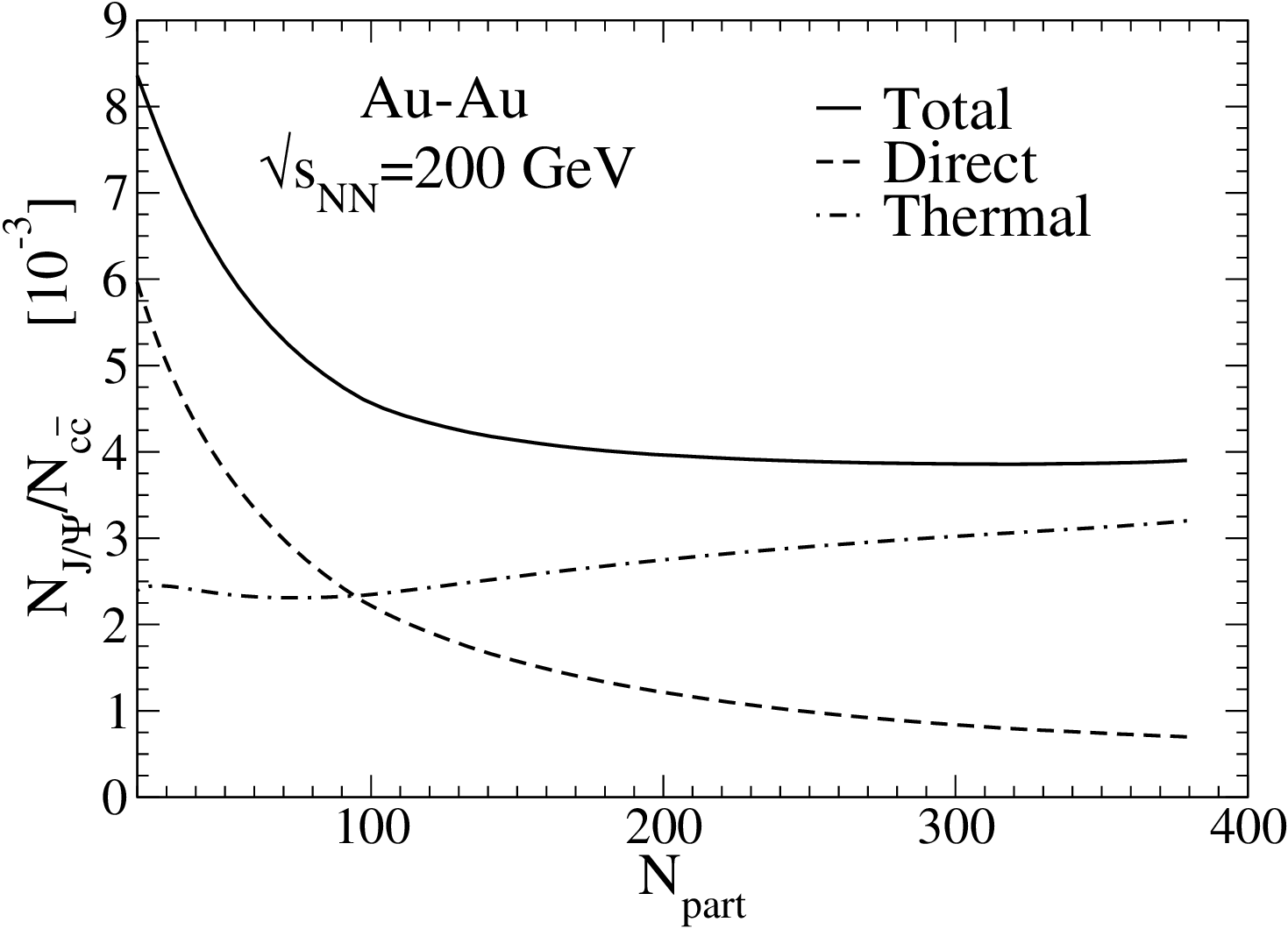,width=0.48\textwidth,clip=}}}
\caption{Comparison of the centrality dependence of the
  $N_{J/\Psi}/N_{c\bar{c}}$ ratio at SPS and RHIC full energy (full line). 
The direct (thermal) contributions are shown separately, respectively in
  the dashed (dot-dashed) lines.}
\label{fig:j_ncc}
\end{figure}

Since the phenomenological extrapolations for absolute numbers of 
primordial $J/\Psi$ and charm-quark production up to RHIC energies,
which are input parameters to our calculations, 
are beset with appreciable uncertainties, it is desirable   
to define a quantity which reduces this sensitivity. 
Therefore, we show in Fig.~\ref{fig:j_ncc} the predictions of our
two-component approach for centrality dependencies of the ratio 
$N_{J/\Psi}/N_{c\bar{c}}$, which, in anticipation of open-charm 
measurements at RHIC, also has the virtue of being composed of  
experimental observables\footnote{However, this ratio may still be
sensitive to, \eg,  shadowing effects: since the \jj\ 
yield, being mostly thermal, goes as $N_{c\bar{c}}^2$ a modification 
of $c\bar{c}$ production due to nuclear
shadowing does not cancel out in this ratio.}. 
Whereas at SPS energies this quantity exhibits a monotonous 
decrease with increasing number of participants (left panel of
Fig.~\ref{fig:j_ncc}), it saturates already for rather peripheral 
collisions at full RHIC energy. Note that the decrease of the thermal
component for $N_{part}\ge100$ at SPS is caused by canonical ensemble
effects, while at RHIC statistical production, for the most part,  
proceeds in the grand-canonical limit entailing a smooth increase 
with centrality. 
Our approach thus clearly discriminates between standard $J/\Psi$ 
suppression as opposed to thermal regeneration at full RHIC 
energy.   

\begin{figure}[!h]
\includegraphics[width=0.5\textwidth,clip=]{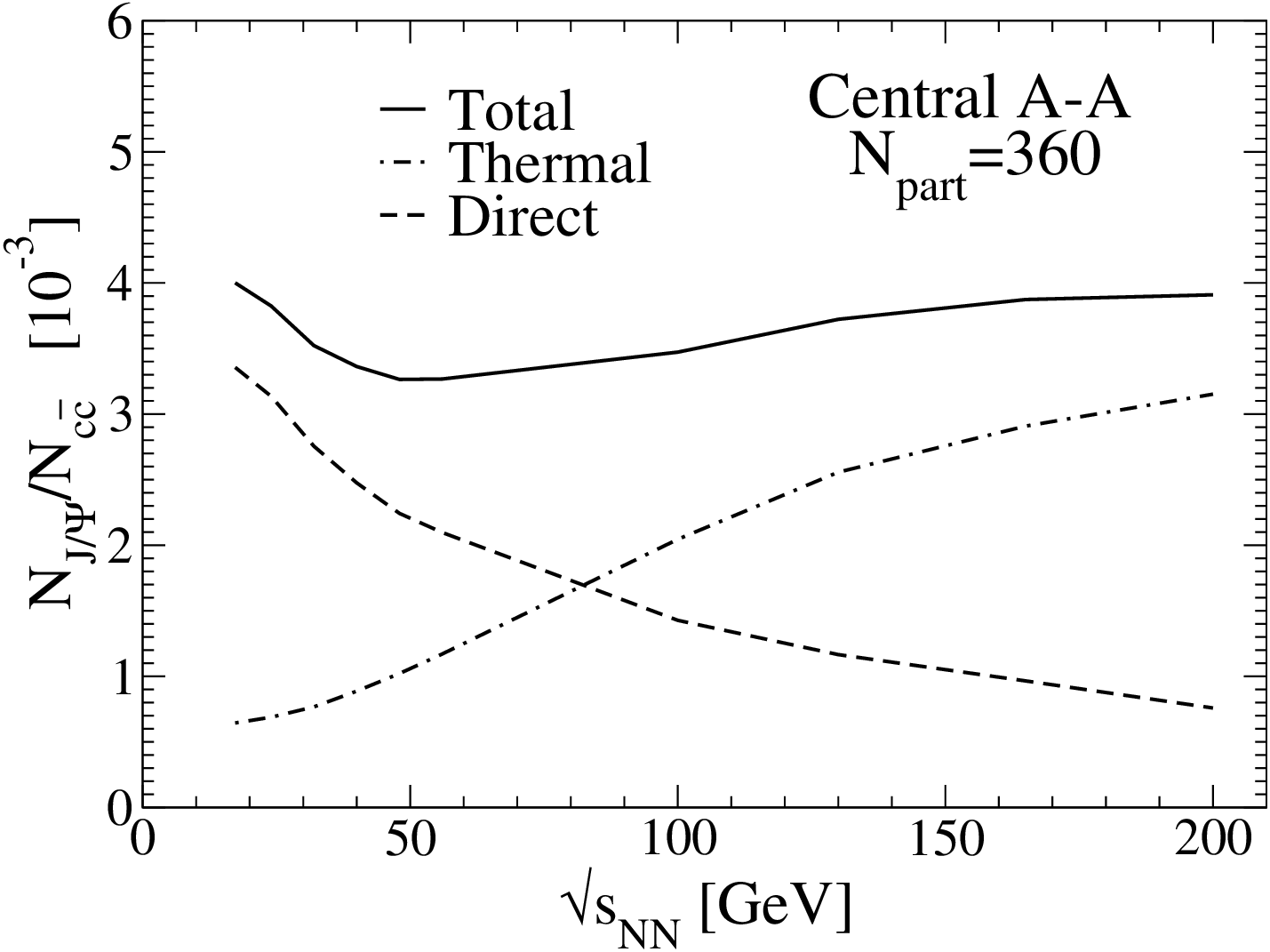} 
\caption{Excitation function of the $N_{J/\Psi}/N_{c\bar{c}}$ ratio
    (full line). The interplay between the direct contribution (dashed
    line) and the thermal component (dot-dashed line) results in a
    minimum in the excitation function around $\sqrt{s_{NN}}\simeq 40$
    GeV.}
\label{fig:excitation}
\end{figure}

It is therefore important (and experimentally feasible at RHIC) 
to map out the transition between
the regimes of predominantly direct to thermal 
charmonium production, as has been first pointed out in Ref.~\cite{GR01}.
In Fig.~\ref{fig:excitation} we present an updated prediction
of the excitation function for $N_{J/\Psi}/N_{c\bar{c}}$ 
ratio\footnote{As compared to our earlier results, we here also 
incorporate the corrections from feeddown of excited charmonia 
and from hadronic suppression. Within the uncertainty band quoted 
in Ref.~\cite{GR01}, the results agree.}.  
The ratio exhibits a nontrivial minimum structure around 
$\sqrt{s_{NN}} \simeq 40$~GeV, which is a marked feature
of the interplay between hard and thermal production (assuming
no anomalies in open-charm production, cf.~Ref.~\cite{Go00}). 

For practical purposes, it is also of interest to convert our results 
into an absolute number of \jj's predicted for RHIC. Keeping in mind 
the uncertainties mentioned above, we find that the combined 
statistical and direct \jj\ yield amounts to rapidity density 
$dN/dy \sim 1$x$10^{-2}$ for central collisions at full RHIC energy.     


\section{Comparison to other works}
\label{sec:comp}

The evaluation of $J/\Psi$ dissolution in a QGP dates back to the 
late 70's / early 80's and has been discussed in many facets.
In our approach, we have combined an in-medium reduced binding energy with
parton-induced destruction, which we have treated in quasifree 
approximation due to small (if any) binding energies of the 
various charmonium states. The latter feature is also the main origin
of the rather strong suppression we find using rather moderate 
values of the strong coupling constant, $\alpha_s\simeq0.3$, which,
of course, is to be regarded as an effective parameter which we adjusted 
to describe the NA50 data. Due to a $\sim$20\% contribution from 
statistical $J/\Psi$ production at SPS energies, our QGP suppression
required to reproduce the NA50 data is stronger than in calculations 
based on the suppression effect alone. On the other hand, within the 
quasifree approximation, the discrimination between different charmonia
is less pronounced than within ``threshold''-type approaches where higher 
charmonium states are completely suppressed once the QGP temperature 
exceeds the relevant value.    

Concerning the evaluation of thermal (statistical) recombination,  
our approach parallels the one initiated in Refs.~\cite{BS00a,GKSG01}, 
which is based on an open-charm abundance generated by hard
production, available for recombination at $T_c$. The main difference
to these analyses is that we do not invoke an enhancement of
open-charm production beyond standard extrapolations of $N$-$N$
collisions (also, we include corrections from incomplete thermalization
of charm quarks in relaxation time approximation).
 Such an enhancement is needed if one aims at describing
the NA50 data in terms of thermal production alone (assuming that
directly produced charmonia have been completely dissolved). 
The NA50 suppression pattern towards more peripheral collisions
then emerges as a canonical ensemble effect, requiring ``anomalous'' 
open-charm enhancement factors of around $\sim 5$~\cite{KG01}. 
Theoretically, such an increase in hard production in a heavy-ion
environment is not easily justified. Also, the NA38 findings
on the intermediate-mass dimuon excess~\cite{na38-00a} allow for a maximal
open-charm enhancement of a factor of $\sim$3 for central collisions 
(less for peripheral)\footnote{In our picture, this excess is attributed
to thermal radiation~\cite{RS00,GKP00}.}. Furthermore, imposing 
statistical production on small-size systems (or even $p$-$A$ reactions) 
eventually contradicts nuclear absorption systematics.  
  
Finally we refer to Ref.~\cite{TSR00}, where the $J/\Psi$ abundance 
in an expanding QGP has been evaluated using rate equations with 
gluon photodissociation and its reverse reaction as microscopic input.  
The space-time evolution in ref.~\cite{TSR00} occurs entirely 
within the QGP phase, and 
in-medium modifications to the $J/\Psi$ binding energy have not been
applied. Consequently, the time evolution of \jj\ formation in that 
approach is quite different from our results, leaning towards 
early production~\cite{TSR01}, which also entails substantially
larger results for the final $N_{J/\Psi}/N_{c\bar{c}}$ ratio, 
reaching values of up to $20$-$30$x$10^{-3}$ for central 
$Au$-$Au$ at $\sqrt{s}_{NN}$=200~GeV. This is a factor of 5-8
larger than our estimates, cf.~Fig.~\ref{fig:excitation}.

\section{Conclusions and outlook}
\label{sec:concl}
In summary, we have developed a two-component model for charmonium 
production in heavy-ion collisions within a comprehensive thermal 
evolution scenario. The two sources of charmonium production are 
(i) ``direct'' \jj's, arising from primordial $N$-$N$ collisions, 
subjected to nuclear, QGP and hadronic suppression, as well as  (ii) 
statistical recombination of $c$ and $\bar c$ quarks at the
hadronization transition subjected to hadronic dissociation only.  
The QGP dissociation of contribution (i) has been evaluated using
in-medium charmonium binding energies which led us to introduce  
``quasifree'' destruction as the dominant suppression 
mechanism. Inelastic hadronic interactions have been
estimated within earlier proposed effective lagrangian approaches,
and turned out to give very moderate corrections (up to 10\% 
for excited charmonium states). Contribution (ii) has been based on
open-charm abundances as inferred from $N$-$N$ collisions (without 
``anomalous'' enhancement), and incomplete thermalization has been 
incorporated via a relaxation time approximation.      
Taken together, with an effective strong coupling constant as a single
parameter, the measured centrality dependence of $J/\Psi$ production 
at the SPS can be reasonably well described. 
A potential discrepancy has been identified in the $\Psi'/\Psi$ ratio, 
which we believe to reside in shortcomings of the  
calculations for hadronic $\Psi'$ dissociation.
The latter is difficult to assess in purely hadronic models and 
might well be underestimated. 

We have extrapolated our approach to higher energies. The pertinent 
excitation function for the $N_{J/\Psi}/N_{c\bar{c}}$ ratio in central
collisions exhibits a non-trivial minimum structure around 
$\sqrt{s_{NN}} \sim 40$~GeV signalling the transition from (predominantly)
direct to thermal production~\cite{GR01}. For the centrality dependence 
of this ratio at full RHIC energy, we predict a rather flat behavior for
 participant 
numbers beyond $N_{part}\simeq 150$. The absolute number of produced 
$J/\Psi$ mesons in central collisions at RHIC turns out to be close
to what one expects from nuclear absorption alone, \ie, an  
80\% QGP suppression is essentially regenerated by thermal production at 
hadronization.   

We emphasize again that the present analysis should also be considered 
as another step towards establishing a coherent picture of 
high-energy heavy-ion collisions as proceeding through a thermal evolution
including QGP formation. Charmonium production has been linked with   
other observables such as hadro-chemistry~\cite{pbm96-99,Beca}, dilepton 
production at low and intermediate mass~\cite{RW99,Ra01,RS00}, etc. 

Concerning directions for future work, it will be necessary
to reiterate the question of in-medium effects in hadronic
dissociation cross sections especially for the lightly 
bound charmonium states.
Here, quark-exchange models appear to be better
suited than hadronic ones, allowing for modifications of 
effective confining potentials and
constituent quark masses~\cite{BSW01} -- 
possibly related to effects of chiral restoration -- 
on a microscopic level.
Furthermore, the NA50 data on \jj\ transverse momentum 
distributions~\cite{na50-01a} need to be addressed as  
another test of our approach.
It is also mandatory to improve on our schematic treatment of
the charm-quark thermalization. Here, valuable
information can be expected from experiment in terms of both
single- and di-lepton spectra from semileptonic open-charm 
decays~\cite{Shu97,GKP98}.

\begin{acknowledgments}
We thank E.V. Shuryak for his interest throughout the course of this
work. We are grateful to C.-Y. Wong and T. Barnes for kindly sharing 
their results with us and for interesting discussions. 
We furthermore thank G. Chanfray and K. Haglin for useful communications. 
This work was supported by the U.S. Department of Energy under Grant
No. DE-FG02-88ER40388. 
\end{acknowledgments}



\end{document}